\documentclass[12pt,a4paper]{article}
\pdfoutput=1
\usepackage{graphicx}
\usepackage[T1]{fontenc}
\usepackage[utf8]{inputenc}
\usepackage{textcomp}
\usepackage[sc,osf]{mathpazo}
\usepackage{a4wide}  
\usepackage{latexsym,amsthm,amsfonts,amsmath,mathrsfs,amssymb}
\usepackage{booktabs} % for much better looking tables
\usepackage[unicode,implicit]{hyperref}
\hypersetup{%
  pdftitle    = {alpha'-corrected four-dimensional black holes in string theory}
  pdfkeywords = {Yang-Mills, instanton, black hole, string theory, entropy,
    branes, supergravity, supersymmetry, corrections},
  pdfauthor   = {Pablo Cano, Samuele Chimento, Patrick Meessen, Tomas Ortin,
    Pedro F. Ramirez and Alejandro Ruiperez},
  plainpages  = true,
  colorlinks  = true,
  citecolor   = blue,
  urlcolor    = red,
  linkcolor   = black
}
\newcommand{\hepth}[1]{{\tt
\href{http://www.arXiv.org/abs/hep-th/#1}{hep-th/#1}}}
\newcommand{\grqc}[1]{{\tt
\href{http://www.arXiv.org/abs/gr-qc/#1}{gr-qc/#1}}}

\newcommand{\arxiv}[1]{{\tt arXiv:\href{http://www.arXiv.org/abs/#1}{#1}}}
\newcommand{\doi}[1]{{\tt DOI:\href{http://dx.doi.org/#1}{#1}}}

\usepackage{tikz}\newcommand{\FPAUO}[2]{
\tikz[scale=.13,
         Uniovi/.style={color=green!51!blue, fill=green!51!blue}
 ] {
 \fill[Uniovi] (0,0) circle (10);
 \fill[white] (0,7) circle (1.5);
 \draw[Uniovi] (-2,7.5) rectangle (2,5.5);
 \fill[white] (-0.3,6.6) rectangle (0.3,0);   % 1.7 cm 
 \fill[white] ( -0.9,6.2) rectangle (.9 ,5.6);
 \fill[white] (-1.4, 5.2) rectangle (1.4, 4.6);
 \fill[white] (0,0) ellipse (3.5 and 4);
 \fill[Uniovi] (-2.5,0.3) rectangle (2.5,-0.3);
 \fill[Uniovi] (-2,2.3) rectangle (2,1.7);
 \fill[Uniovi] (-2,-2.3) rectangle (2,-1.7);
 \fill[white] (-4.5,5.5) rectangle (-2.7,4.9);
 \fill[white] (-3.9,6.1) rectangle (-3.3,4.3);
 \fill[white] (4.5,5.5) rectangle (2.7,4.9);
 \fill[white] (3.9,6.1) rectangle (3.3,4.3);
 \foreach \x in { 0,..., 3 }
   \foreach \y in { 0,...,\x}
    {
     \fill[white] (-6-\x*0.7+\y*1.4,3.5-\x *1.97) -- (-5.6-\x*0.7+\y*1.4,2.4-\x *1.97) -- (-6.4-\x*0.7+\y*1.4,2.4-\x *1.97) -- cycle;
     \fill[white] (6-\x*0.7+\y*1.4,3.5-\x *1.97) -- (5.6-\x*0.7+\y*1.4,2.4-\x *1.97) -- (6.4-\x*0.7+\y*1.4,2.4-\x *1.97) -- cycle;
   };
 \draw (0,-6) node[
                               text centered, 
                               color=white, 
                               font={\fontsize{8}{4}\sffamily\selectfont}
                             ] {FPAUO-#1/#2};
}} 
\makeatletter
\@addtoreset{equation}{section}
\makeatother

\begin{document}

\begin{flushright}
\small
\FPAUO{18}{07}\\
IFT-UAM/CSIC-17-091\\
IFUM-1061-FT\\
%\texttt{arXiv:yymm.nnnnn [hep-th]}\\
August 9\textsuperscript{th}, 2018\\
\normalsize
\end{flushright}

\begin{center}

{\Large {\bf {Beyond the near-horizon limit: Stringy corrections\\[.5cm] 
to Heterotic Black Holes }}}

\vspace{1.5cm}

\renewcommand{\thefootnote}{\alph{footnote}}
{\sl\large Pablo A.~Cano$^{1}$}\footnote{\tt pablo.cano[at]uam.es, samuele.chimento[at]csic.es, meessenpatrick[at]uniovi.es, tomas.ortin[at]csic.es, ramirez.pedro[at]mi.infn.it, alejandro.ruiperez[at]uam.es}, 
{\sl\large Samuele Chimento$^{1}$},
{\sl\large Patrick Meessen$^{2}$},
{\sl\large Tom\'{a}s Ort\'{\i}n$^{1}$},\\[.5cm]
{\sl\large Pedro F.~Ram\'{\i}rez$^{3}$}
{\sl\large and~Alejandro~Ruip\'erez$^{1}$}

\setcounter{footnote}{0}
\renewcommand{\thefootnote}{\arabic{footnote}}

\vspace{0.5cm}

${}^{1}${\it Instituto de F\'{\i}sica Te\'orica UAM/CSIC\\
C/ Nicol\'as Cabrera, 13--15,  C.U.~Cantoblanco, E-28049 Madrid, Spain}\\ 
\vspace{0.2cm}

${}^{2}${\it HEP Theory Group, Departamento de F\'{\i}sica, Universidad de Oviedo\\
  Avda.~Calvo Sotelo s/n, E-33007 Oviedo, Spain}\\

\vspace{0.2cm}

${}^{3}${\it INFN, Sezione di Milano, Via Celoria 16, 20133 Milano, Italy}

\vspace{.8cm}

%%%%%%%%%%%%%%%%%%%%%%%%%%%%%%%%%%%%%%%%%%%%%%%%%%%%%%%%%%%%%%%%%%%%%%

{\bf Abstract}
\end{center}
\begin{quotation}
  {\small We study the first-order in $\alpha'$ corrections to 4-charge black
    holes (with the Reissner-Nordstr\"om black hole as a particular example)
    beyond the near-horizon limit in the Heterotic Superstring effective
    action framework. The higher-curvature terms behave as delocalized sources
    in the equations of motion and in the Bianchi identity of the 3-form. For some
    charges, this introduces a shift between their values measured at the
    horizon and asymptotically. Some of these corrections and their associated
    charge shifts, but not all of them, can be canceled using appropriate
    SU$(2)$ instantons for the heterotic gauge fields. The entropy, computed
    using Wald's formula, is in agreement with the result obtained via
    microstate counting when the delocalized sources are properly taken into
    account.}
\end{quotation}

\newpage
%%%%%%%%%%%%%%%%%%%%%%%%%%%%%%%%%%%%%%%%%%%%%%%%%%%%%%%%%%%%%%%%%%%%%%
%%%%%%%%%%%%%%%%%%%%%%%%%%%%%%%%%%%%%%%%%%%%%%%%%%%%%%%%%%%%%%%%%%%%%%
%%%%%%%%%%%%%%%%%%%%%%%%%%%%%%%%%%%%%%%%%%%%%%%%%%%%%%%%%%%%%%%%%%%%%%
%%%%%%%%%%%%%%%%%%%%%%%%%%%%%%%%%%%%%%%%%%%%%%%%%%%%%%%%%%%%%%%%%%%%%%
\pagestyle{plain}
%%%%%%%%%%%%%%%%%%%%%%%%%%%%%%%%%%%%%%%%%%%%%%%%%%%%%%%%%%%%%%%%%%%%%%
%%%%%%%%%%%%%%%%%%%%%%%%%%%%%%%%%%%%%%%%%%%%%%%%%%%%%%%%%%%%%%%%%%%%%%
%%%%%%%%%%%%%%%%%%%%%%%%%%%%%%%%%%%%%%%%%%%%%%%%%%%%%%%%%%%%%%%%%%%%%%
%%%%%%%%%%%%%%%%%%%%%%%%%%%%%%%%%%%%%%%%%%%%%%%%%%%%%%%%%%%%%%%%%%%%%%

\tableofcontents

%\newpage

%%%%%%%%%%%%%%%%%%%%%%%%%%%%%%%%%%%%%%%%%%%%%%%%%%%%%%%%%%%%%%%%%%%%%%
%%%%%%%%%%%%%%%%%%%%%%%%%%%%%%%%%%%%%%%%%%%%%%%%%%%%%%%%%%%%%%%%%%%%%%
%%%%%%%%%%%%%%%%%%%%%%%%%%%%%%%%%%%%%%%%%%%%%%%%%%%%%%%%%%%%%%%%%%%%%%
%%%%%%%%%%%%%%%%%%%%%%%%%%%%%%%%%%%%%%%%%%%%%%%%%%%%%%%%%%%%%%%%%%%%%%
\section{Introduction}
%%%%%%%%%%%%%%%%%%%%%%%%%%%%%%%%%%%%%%%%%%%%%%%%%%%%%%%%%%%%%%%%%%%%%%
%%%%%%%%%%%%%%%%%%%%%%%%%%%%%%%%%%%%%%%%%%%%%%%%%%%%%%%%%%%%%%%%%%%%%%
%%%%%%%%%%%%%%%%%%%%%%%%%%%%%%%%%%%%%%%%%%%%%%%%%%%%%%%%%%%%%%%%%%%%%%
%%%%%%%%%%%%%%%%%%%%%%%%%%%%%%%%%%%%%%%%%%%%%%%%%%%%%%%%%%%%%%%%%%%%%%

In our recent work Ref.~\cite{Chimento:2018kop} we have constructed a large
family of solutions of the Heterotic Superstring effective action to first
order in $\alpha'$. Generically, these solutions describe well-known systems
consisting of intersections of fundamental strings (F1) with momentum flowing
along them (W), solitonic 5-branes (NS or S5) and Kaluza-Klein monopoles
(KK).  The 5-dimensional, extremal, 3-charge black holes studied in
Refs.~\cite{Cano:2017qrq,Cano:2018qev} are simple members of this family with
no KK monopoles and they describe the first-order in $\alpha'$ corrections of
the heterotic version of the Strominger-Vafa black hole
\cite{Strominger:1996sh}.  Our main task in this paper will be to study the
case with KK monopoles. The corresponding solutions are 4-dimensional,
extremal, 4-charge black holes which will contain the first-order in $\alpha'$
corrections to the heterotic version of the black holes whose microscopic
entropy was computed and compared with the supergravity result in
Refs.~\cite{Maldacena:1996gb,Johnson:1996ga,Maldacena:1996ky}.\footnote{See
  also Refs.~\cite{Kraus:2006wn,Sen:2007qy} and references therein. The
  $\alpha' \rightarrow 0$ limits of these solutions are well known and were
  first obtained in Ref.~\cite{Cvetic:1995uj} directly in the heterotic
  version.}

The agreement between the Bekenstein-Hawking (BH) entropy of 4- and
5-dimensional black holes and the degeneracy of string microstates in the
backgrounds mentioned above, initially obtained at the $\alpha' \rightarrow 0$
level in regimes in which the $\alpha'$ corrections can be safely ignored, is
one of the triumphs of String Theory. These results have been extended in
several directions to include rotation \cite{Breckenridge:1996is}, non-trivial
topology of the horizon (black rings) \cite{Elvang:2004ds} etc.

A very important question to study is whether this agreement between the
values of the BH entropy calculated by macroscopic and microscopic methods still
holds when $\alpha'$ corrections (genuinely stringy effects associated to the
finite string size $\ell_{S}=\sqrt{\alpha'}$) are taken into account.

In the calculation of the BH entropy by microstate counting the AdS/CFT
correspondence tools have proven extraordinarily useful, shedding results that
account for all the contributions in the $\alpha'$ perturbative expansion in
the large charge regime.

The near-horizon geometry of all the black hole solutions we consider is
$\mathrm{AdS}_{3} \times \mathrm{S}^{3}/\mathbb{Z}_{n} \times
\mathrm{T}^{4}$. The $\mathrm{AdS}_{3}$ and $\mathrm{S}^{3}$ factors are
standard in the three-charge family of extremal black holes, and the quotient
of the sphere by $\mathbb{Z}_{n}$ is related to the presence of a charge-$n$
KK monopole. Heterotic String Theory on this background was studied in
Ref.~\cite{Kutasov:1998zh}, identifying the central charges of the dual
CFT. Then, applying the Cardy formula one obtains the following expression for
the entropy

\begin{equation}
\label{eq:entropyexact}
\mathbb{S}_{\rm CFT}=2\pi \sqrt{N_{F1} N_{W} \left( k+2 \right)} \, ,
\end{equation}

\noindent
where $N_{F1}$ is the number of fundamental strings present in the background,
$N_{W}$ is the number of units of momentum flowing along them and $k$ is the
total level of affine algebra $\widehat{\mathrm{SL}(2)}$ in the right-moving
sector. This number, minus two units, was identified in
Ref.~\cite{Kutasov:1998zh} with the product of the KK monopole charge and S5
charge: $k=n \mathcal{Q}_{S5}+2$. As we will see along this
paper and in the discussion section, distinguishing correctly between charges
(total charges, evaluated at spatial infinity) and numbers of stringy objects
(KK monopoles and S5-branes in this case) is essential in order to compare
this microscopic result with the macroscopic one.

The macroscopic (``supergravity'') calculation of the $\alpha'$ corrections to
the BH entropy faces a number of difficulties:

\begin{enumerate}
\item Finding the $\alpha'$-corrected solutions is a very complicated task,
  owing to the higher-order in curvature terms present in the equations of
  motion and the complicated interactions between them. In the Heterotic
  Superstring effective action there is an infinite series of terms related to
  the supersymmetrization of the Chern-simons terms present in the NS 3-form
  that can be introduced in an interative way following
  Refs.~\cite{Bergshoeff:1988nn,Bergshoeff:1989de} but there are further terms
  of higher order in the curvature that seem to be unrelated to them first
  appearing at $\mathcal{O}(\alpha'^{3})$ \cite{Kehagias:1997cq}.
\item The BH entropy of the $\alpha'$-corrected solutions is no longer simply
  given by the area of the horizon, and one needs to use the Wald formula
  \cite{Wald:1993nt,Iyer:1994ys}.
\end{enumerate}

In order to circumvent these difficulties one may try to work with the
near-horizon region of the solution only, assuming that it will have the same
geometry after the $\alpha'$ corrections are taken into acount.  The entropy
function formalism developed by Sen \cite{Sen:2005iz,Sen:2007qy} provides an
elegant and powerful strategy to find the near-horizon solutions of extremal
black holes and to compute their entropy, making a comparison with the
microscopic result Eq.~(\ref{eq:entropyexact}) possible. This approach has
important drawbacks, though: it is not guaranteed that a solution
interpolating between the near-horizon geometry and Minkowski spacetime
describing an asymptotically-flat black-hole spacetime exists and, if it does,
it does not give any information on how the non-linear interactions introduced
by the higher-order corrections affect the physical properties of the
solution, such as the values of the conserved charges.
 
The family of solutions constructed in Ref.~\cite{Chimento:2018kop} makes
unnecessary the restriction to the near-horizon limit, because they can
describe the complete black hole spacetime to first order in $\alpha'$, as
shown for the 5-dimensional case in Ref.~\cite{Cano:2018qev}. This allows us
to take into account the non-linear interactions and compute explicitly the
asymptotic charges. In the S5 and W cases, these will receive contributions
from localized sources signaling the presence of the corresponding fundamental
objects in the String Theory background, and contributions from the non-linear
interactions that arise at first order in $\alpha'$. Being able to make this
distinction is essential in order to write the BH entropy in terms of the same
variables used in the counting of string microstates.

The non-linear contributions to the total S5 charge are analogous to those of
SU$(2)$ instantons over the KK monopole\footnote{When the charge of the KK
  monopole is larger than one, the Yang-Mills instantons are somewhat exotic
  because the KK monopole contains a conical singularity. However, this
  singularity is resolved in the full supergravity metric by a conformal
  factor and the corresponding instantons are regular in 10-dimensional
  spacetime as well. Nevertheless, this deficit in the angle is reflected in
  terms of a fractional (although discrete) instanton number.} with the wrong
sign and can be exactly cancelled through the introduction of heterotic
SU$(2)$ gauge fields with the same instanton configuration.\footnote{These
  Yang-Mills fields coincide with those of some of the non-Abelian
  supergravity solutions we have studied in
  Refs.~\cite{Cano:2017qrq,Meessen:2008kb,Meessen:2015enl,Ramirez:2016tqc,Cano:2017sqy,Avila:2017pwi}.}
It can, then, be argued that certain components of the fields associated to
the S5 charge will not receive any further $\alpha'$ corrections.

The non-linear contributions to the momentum (W), though, are of a more
mysterious nature. They take the form of the contribution of a generalization
of the the electric sector of the dyon of Ref.~\cite{Ramirez:2016tqc} with the
standard sign and cannot be cancelled using the same Green-Schwarz-type
mechanism. However, it can still be argued that further $\alpha'$ corrections
connected to the supersymmetrization of the Chern-Simons terms will vanish or
will be arbitrarily small.

%As remarked in Ref.~\cite{Cano:2018qev}, this correction occurs in $uu$
%component of the Einstein equation as a negative contribution to the
%corresponding component of the energy-momentum tensor. This contribution is
%similar to the one considered in Ref.~\cite{Maldacena:2018gjk} which gives
%rise to a 4-dimensional traversable wormhole. In that case, the origin of the
%negative contribution to the energy-momentum tensor is quantum-mechanical. In
%the case at hands, $\alpha'$ corrections are not, strictly speaking, quantum
%mechanical, but end up playing a similar r\^ole, giving rise to classically
%forbidden gravitational fields such as the globally regular black hole of
%Ref.~\cite{Cano:2018aod}.

Having the 4-dimensional, extremal, 4-charge black holes with their
first-order in $\alpha'$ corrections under control just leaves us with the
calculation of the entropy using Wald's formula. This calculation can be
conveniently performed directly in 10 dimensions using the same trick we used
in the 5-dimensional case Ref.~\cite{Cano:2018qev} and the result, to first
order in $\alpha'$, is found to be

\begin{equation}
\label{eq:entropy}
\mathbb{S}_{\rm Wald}
=
2\pi \sqrt{N_{F1} N_{W} n N_{S5}} \left( 1+ \frac{1}{n N_{S5}} +\cdots\right)\, .
\end{equation}

\noindent
Comparing this macroscopic result with the expansion in the large charge limit
of the result coming from the counting of string microstates
Eq.~(\ref{eq:entropyexact}) we see that they agree with each other upon the
identification $k=n N_{S5}$, where $N_{S5}$ is the number of S5-branes. This
is the main result of this article.

This article is organized as follows: the Heterotic Superstring effective
action at first order in $\alpha'$ is described in
Section~\ref{sec-heteroticalpha}. In Section~\ref{sec-solutions} we review the
generic structure of the 10-dimensional fields for the family of solutions
studied and discuss some aspects of their lower-dimensional descendants. In
Section~\ref{Ingredients} we present the construction of two families of
spherically symmetric selfdual instantons over KK monopoles, which can be used
as gauge fields in the solution. Additional details about these constructions
are contained in Appendix~\ref{sec:regular_inst}. Our most important results
are described in Section~\ref{sec-all}: the description of the 4-charge
corrected black hole and the identification of the parameters in terms of
fundamental objects in the String Theory background. In Section~\ref{sec-4d}
we briefly describe some physical properties of our solutions from an
effective 4-dimensional perspective. We particularize the discussion for the
special example of an extremal Reissner-Nordstr\"om solution and comment on
the non-perturbative nature of the small black holes. In
Section~\ref{sec-entropy} we compute in detail the Wald entropy.  Finally, we
discuss our results and compare them with the previous literature in
Section~\ref{sec-discussion}.

%%%%%%%%%%%%%%%%%%%%%%%%%%%%%%%%%%%%%%%%%%%%%%%%%%%%%%%%%%%%%%%%%%%%%%
%%%%%%%%%%%%%%%%%%%%%%%%%%%%%%%%%%%%%%%%%%%%%%%%%%%%%%%%%%%%%%%%%%%%%%
%%%%%%%%%%%%%%%%%%%%%%%%%%%%%%%%%%%%%%%%%%%%%%%%%%%%%%%%%%%%%%%%%%%%%%
%%%%%%%%%%%%%%%%%%%%%%%%%%%%%%%%%%%%%%%%%%%%%%%%%%%%%%%%%%%%%%%%%%%%%%
\section{The Heterotic Superstring effective action to
  \texorpdfstring{$\mathcal{O}(\alpha')$}{O(α')}}
\label{sec-heteroticalpha}
%%%%%%%%%%%%%%%%%%%%%%%%%%%%%%%%%%%%%%%%%%%%%%%%%%%%%%%%%%%%%%%%%%%%%%
%%%%%%%%%%%%%%%%%%%%%%%%%%%%%%%%%%%%%%%%%%%%%%%%%%%%%%%%%%%%%%%%%%%%%%
%%%%%%%%%%%%%%%%%%%%%%%%%%%%%%%%%%%%%%%%%%%%%%%%%%%%%%%%%%%%%%%%%%%%%%
%%%%%%%%%%%%%%%%%%%%%%%%%%%%%%%%%%%%%%%%%%%%%%%%%%%%%%%%%%%%%%%%%%%%%%

In order to describe the Heterotic Superstring effective action to
$\mathcal{O}(\alpha')$ as given in Ref.~\cite{Bergshoeff:1989de} (but in
the string frame), we start by defining the zeroth-order 3-form field strength of
the Kalb-Ramond 2-form $B$:

\begin{equation}
H^{(0)} \equiv dB\, ,  
\end{equation}

\noindent
and constructing with it the zeroth-order torsionful spin connections

\begin{equation}
{\Omega}^{(0)}_{(\pm)}{}^{{a}}{}_{{b}} 
=
{\omega}^{{a}}{}_{{b}}
\pm
\tfrac{1}{2}{H}^{(0)}_{{\mu}}{}^{{a}}{}_{{b}}dx^{{\mu}}\, ,
\end{equation}

\noindent
where ${\omega}^{{a}}{}_{{b}}$ is the Levi-Civita spin connection
1-form.\footnote{We follow the conventions of Ref.~\cite{Ortin:2015hya} for
  the spin connection and the curvature.}  With them we
define the zeroth-order Lorentz curvature 2-form and Chern-Simons 3-forms

\begin{eqnarray}
{R}^{(0)}_{(\pm)}{}^{{a}}{}_{{b}}
& = & 
d {\Omega}^{(0)}_{(\pm)}{}^{{a}}{}_{{b}}
- {\Omega}^{(0)}_{(\pm)}{}^{{a}}{}_{{c}}
\wedge  
{\Omega}^{(0)}_{(\pm)}{}^{{c}}{}_{{b}}\, ,
\\
& & \nonumber \\
{\omega}^{{\rm L}\, (0)}_{(\pm)}
& = &  
d{\Omega}^{ (0)}_{(\pm)}{}^{{a}}{}_{{b}} \wedge 
{\Omega}^{ (0)}_{(\pm)}{}^{{b}}{}_{{a}} 
-\tfrac{2}{3}
{\Omega}^{ (0)}_{(\pm)}{}^{{a}}{}_{{b}} \wedge 
{\Omega}^{ (0)}_{(\pm)}{}^{{b}}{}_{{c}} \wedge
{\Omega}^{ (0)}_{(\pm)}{}^{{c}}{}_{{a}}\, .  
\end{eqnarray}

Next, we introduce the gauge fields. We will only activate a
$\mathrm{SU}(2)\times \mathrm{SU}(2)$ subgroup of the full gauge group
of the Heterotic Theory and we will denote by
$A^{A_{1,2}}$ ($A_{1,2}=1,2,3$) the components. The gauge field strength and
the Chern-Simons 3-form of each $\mathrm{SU}(2)$ factor are defined by

\begin{eqnarray}
{F}^{A}
& = & 
d{A}^{A}+\tfrac{1}{2}\epsilon^{ABC}{A}^{B}\wedge{A}^{C}\, , 
\\
& & \nonumber \\
{\omega}^{\rm YM}
& = & 
dA^{A}\wedge {A}^{A}
+\tfrac{1}{3}\epsilon^{ABC}{A}^{A}\wedge{A}^{B}\wedge{A}^{C}\, .
\end{eqnarray}

Then, we are ready to define recursively 

\begin{eqnarray}
H^{(1)}
& = & 
d{B}
+\frac{\alpha'}{4}\left({\omega}^{\rm YM}
+{\omega}^{{\rm L}\, (0)}_{(-)}\right)\, ,  
\nonumber \\
& & \nonumber \\
{\Omega}^{(1)}_{(\pm)}{}^{{a}}{}_{{b}} 
& = & 
{\omega}^{{a}}{}_{{b}}
\pm
\tfrac{1}{2}{H}^{(1)}_{{\mu}}{}^{{a}}{}_{{b}}dx^{{\mu}}\, ,
\nonumber \\
& & \nonumber \\
{R}^{(1)}_{(\pm)}{}^{{a}}{}_{{b}}
& = & 
d {\Omega}^{(1)}_{(\pm)}{}^{{a}}{}_{{b}}
- {\Omega}^{(1)}_{(\pm)}{}^{{a}}{}_{{c}}
\wedge  
{\Omega}^{(1)}_{(\pm)}{}^{{c}}{}_{{b}}\, ,
\nonumber  \\
& & \nonumber \\
{\omega}^{{\rm L}\, (1)}_{(\pm)}
& = & 
d{\Omega}^{(1)}_{(\pm)}{}^{{a}}{}_{{b}} \wedge 
{\Omega}^{(1)}_{(\pm)}{}^{{b}}{}_{{a}} 
-\tfrac{2}{3}
{\Omega}^{(1)}_{(\pm)}{}^{{a}}{}_{{b}} \wedge 
{\Omega}^{(1)}_{(\pm)}{}^{{b}}{}_{{c}} \wedge
{\Omega}^{(1)}_{(\pm)}{}^{{c}}{}_{{a}}\, .  
\nonumber \\
& & \nonumber \\
H^{(2)}
& = &  
d{B}
+\frac{\alpha'}{4}\left({\omega}^{\rm YM}
+{\omega}^{{\rm L}\, (1)}_{(-)}\right)\, ,  
\end{eqnarray}

\noindent
and so on.

In practice only $\Omega^{(0)}_{(\pm)},{R}^{(0)}_{(\pm)}, \omega^{{\rm L}\,
  (0)}_{(\pm)}, H^{(1)}$ will occur to the order we want to work at, but, often,
it is more convenient to work with the higher-order objects ignoring the terms of
higher order in $\alpha'$ when necessary. Thus we will suppress the $(n)$
upper indices from now on.

Finally, we define three ``$T$-tensors'' associated to the $\alpha'$
corrections

\begin{equation}
\label{eq:Ttensors}
\begin{array}{rcl}
{T}^{(4)}
& \equiv &
\dfrac{3\alpha'}{4}\left[
{F}^{A}\wedge{F}^{A}
+
{R}_{(-)}{}^{{a}}{}_{{b}}
\wedge
{R}_{(-)}{}^{{b}}{}_{{a}}
\right]\, ,
\\
& & \\ 
{T}^{(2)}{}_{{\mu}{\nu}}
& \equiv &
\dfrac{\alpha'}{4}\left[
{F}^{A}{}_{{\mu}{\rho}}{F}^{A}{}_{{\nu}}{}^{{\rho}} 
+
{R}_{(-)\, {\mu}{\rho}}{}^{{a}}{}_{{b}}
{R}_{(-)\, {\nu}}{}^{{\rho}\,  {b}}{}_{{a}}
\right]\, ,
\\
& & \\    
{T}^{(0)}
& \equiv &
{T}^{(2)\,\mu}{}_{{\mu}}\, .
\\
\end{array}
\end{equation}

In terms of all these objects, the Heterotic Superstring effective action in
the string frame and to first-order in $\alpha'$ can be written as

\begin{equation}
\label{heterotic}
{S}
=
\frac{g_{s}^{2}}{16\pi G_{N}^{(10)}}
\int d^{10}x\sqrt{|{g}|}\, 
e^{-2{\phi}}\, 
\left\{
{R} 
-4(\partial{\phi})^{2}
+\tfrac{1}{2\cdot 3!}{H}^{2}
-\tfrac{1}{2}T^{(0)}
\right\}\, ,
\end{equation}

\noindent
where $G_{N}^{(10)}$ is the 10-dimensional Newton constant, $\phi$ is the
dilaton field and the vacuum expectation value of $e^{\phi}$ is the Heterotic
Superstring coupling constant $g_{s}$. $R$ is the Ricci scalar of the
string-frame metric $g_{\mu\nu}$.

The derivation of the complete equations of motion is quite a complicated
challenge.  Following Ref.~\cite{Bergshoeff:1992cw}, we separate the
variations with respect to each field into those corresponding to occurrences
via ${\Omega}_{(-)}{}^{{a}}{}_{{b}}$, that we will call \textit{implicit}, and
the rest, that we will call \textit{explicit}:

\begin{eqnarray}
\delta S 
& = &  
\frac{\delta S}{\delta g_{\mu\nu}}\delta g_{\mu\nu}
+\frac{\delta S}{\delta B_{\mu\nu}}\delta B_{\mu\nu}
+\frac{\delta S}{\delta A^{A_{i}}{}_{\mu}}\delta A^{A_{i}}{}_{\mu}
+\frac{\delta S}{\delta \phi} \delta \phi
\nonumber \\
& & \nonumber \\
& = & 
\left.\frac{\delta S}{\delta g_{\mu\nu}}\right|_{\rm exp.}\delta g_{\mu\nu}
+\left.\frac{\delta S}{\delta B_{\mu\nu}}\right|_{\rm exp.}\delta B_{\mu\nu}
+\left.\frac{\delta S}{\delta A^{A_{i}}{}_{\mu}}\right|_{\rm exp.}
\delta A^{A_{i}}{}_{\mu}
+\frac{\delta S}{\delta \phi} \delta \phi
\nonumber \\
& & \nonumber \\
& &
+\frac{\delta S}{ \delta {\Omega}_{(-)}{}^{{a}}{}_{{b}}}
\left(
\frac{\delta {\Omega}_{(-)}{}^{{a}}{}_{{b}}}{\delta g_{\mu\nu}}\delta g_{\mu\nu}
+\frac{\delta {\Omega}_{(-)}{}^{{a}}{}_{{b}}}{\delta B_{\mu\nu}} \delta B_{\mu\nu}
+\frac{\delta {\Omega}_{(-)}{}^{{a}}{}_{{b}}}{\delta A^{A_{i}}{}_{\mu}}\delta
A^{A_{i}}{}_{\mu}
\right)\, .
\end{eqnarray}

We can then apply a lemma proven in Ref.~\cite{Bergshoeff:1989de}: $\delta
S/\delta {\Omega}_{(-)}{}^{{a}}{}_{{b}}$ is proportional to $\alpha'$ and to
the zeroth-order equations of motion of $g_{\mu\nu},B_{\mu\nu}$ and $\phi$
plus terms of higher order in $\alpha'$.

The upshot is that, if we consider field configurations which solve the
zeroth-order equations of motion\footnote{These can be obtained from
  Eqs.~(\ref{eq:eq1})-(\ref{eq:eq4}) by setting $\alpha'=0$. This eliminates
  the Yang-Mills fields, the $T$-tensors and the Chern-Simons terms in $H$.}
up to terms of order $\alpha'$, the contributions to the equations of motion
associated to the implicit variations are at least of second order in
$\alpha'$ and we can safely ignore them here.

If we restrict ourselves to this kind of field configurations, the equations
of motion reduce to 

\begin{eqnarray}
\label{eq:eq1}
R_{\mu\nu} -2\nabla_{\mu}\partial_{\nu}\phi
+\tfrac{1}{4}{H}_{\mu\rho\sigma}{H}_{\nu}{}^{\rho\sigma}
-T^{(2)}{}_{\mu\nu}
& = & 
0\, ,
\\
& & \nonumber \\
\label{eq:eq2}
(\partial \phi)^{2} -\tfrac{1}{2}\nabla^{2}\phi
-\tfrac{1}{4\cdot 3!}{H}^{2}
+\tfrac{1}{8}T^{(0)}
& = &
0\, ,
\\
& & \nonumber \\
\label{eq:eq3}
d\left(e^{-2\phi}\star {H}\right)
& = &
0\, ,
\\
& & \nonumber \\
\label{eq:eq4}
% e^{2\phi}d\left(e^{-2\phi}\star {F}^{A}\right)
% +\epsilon^{ABC}{A}^{B}\wedge \star F^{C}
% +\star {H}\wedge{F}^{A}
% & = & 
% 0\, .
\alpha' e^{2\phi}\mathfrak{D}_{(+)}\left(e^{-2\phi}\star {F}^{A_{i}}\right)
& = & 
0\, ,
\end{eqnarray}

\noindent
where $\mathfrak{D}_{(+)}$ stands for the exterior derivative covariant with
respect to each $\mathrm{SU}(2)$ subgroup and with respect to the torsionful
connection $\Omega_{(+)}$: suppressing the subindices $1,2$ that distinguish
the two subgroups, it takes the explicit form 

\begin{equation}
\label{eq:eq5}
e^{2\phi}d\left(e^{-2\phi}\star {F}^{A}\right)
+\epsilon^{ABC}{A}^{B}\wedge \star F^{C}
+\star {H}\wedge{F}^{A}
= 
0\, .   
\end{equation}

If the ansatz is given in terms of the 3-form field strength, we also need to
solve the Bianchi identity 

\begin{equation}
\label{eq:BianchiH}
d{H}  
-
\tfrac{1}{3}T^{(4)}
=
0\, ,
\end{equation}

\noindent
as well. 

%%%%%%%%%%%%%%%%%%%%%%%%%%%%%%%%%%%%%%%%%%%%%%%%%%%%%%%%%%%%%%%%%%%%%%
%%%%%%%%%%%%%%%%%%%%%%%%%%%%%%%%%%%%%%%%%%%%%%%%%%%%%%%%%%%%%%%%%%%%%%
%%%%%%%%%%%%%%%%%%%%%%%%%%%%%%%%%%%%%%%%%%%%%%%%%%%%%%%%%%%%%%%%%%%%%%
%%%%%%%%%%%%%%%%%%%%%%%%%%%%%%%%%%%%%%%%%%%%%%%%%%%%%%%%%%%%%%%%%%%%%%
\section{The 10-dimensional solutions and their 
\texorpdfstring{$d=5,4$}{d=5,4} descendents}
\label{sec-solutions}
%%%%%%%%%%%%%%%%%%%%%%%%%%%%%%%%%%%%%%%%%%%%%%%%%%%%%%%%%%%%%%%%%%%%%%
%%%%%%%%%%%%%%%%%%%%%%%%%%%%%%%%%%%%%%%%%%%%%%%%%%%%%%%%%%%%%%%%%%%%%%
%%%%%%%%%%%%%%%%%%%%%%%%%%%%%%%%%%%%%%%%%%%%%%%%%%%%%%%%%%%%%%%%%%%%%%
%%%%%%%%%%%%%%%%%%%%%%%%%%%%%%%%%%%%%%%%%%%%%%%%%%%%%%%%%%%%%%%%%%%%%%

The 5- and 4-dimensional black holes we are interested in belong to the class
of $\alpha'$-corrected solutions constructed in
Ref.~\cite{Chimento:2018kop}. These preserve $1/4$ of the 16 supersymmetries
of the heterotic theory and are completely determined by

\begin{enumerate}

\item A choice of 4-dimensional hyperK\"ahler metric

\begin{equation}
\label{eq:metricHK}
d\sigma^{2}=h_{\underline{m}\underline{n}}dx^{m}dx^{n}\, ,
\,\,\,\,\
m,n=\sharp,1,2,3\, ,
\end{equation}

\noindent
with self-dual Riemann curvature 2-form with respect to some standard
orientation.\footnote{We use $\varepsilon^{\sharp 1 2 3 }=+1$ in an
  appropriate Vierbein basis $v^{m}$
\begin{equation}
\label{eq:HKVierbein}
h_{\underline{m}\underline{n}} = v^{p}{}_{\underline{m}}
v^{p}{}_{\underline{n}}\, .
\end{equation}
}If we are interested in 5-dimensional solutions, which are obtained by
compactification on $\mathrm{T}^{5}$, we can use any non-compact 4-dimensional
hyperK\"ahler space. If we are interested in 4-dimensional supersymmetric
solutions, though, the hyperK\"ahler space must admit an additional isometry
and we will take it to have a Gibbons-Hawking metric of the form\footnote{Here
  $\eta=x^{\sharp}$ and we are using the 3-dimensional, curved, indices
  $x,y,z=1,2,3$ which should not be mistaken with coordinates.}

\begin{equation}
\label{eq:GHmetric}
d\sigma^{2}= \mathcal{H}^{-1}(d\eta+\chi)^{2}+\mathcal{H}dx^{x}dx^{x}\, ,
\,\,\,\,\,
d\mathcal{H}
=
\star_{(3)}d\chi\, ,
\end{equation}

\noindent
where $\star_{(3)}$ denotes the Hodge dual in $\mathbb{E}^{3}$.

In the Gibbons-Hawking case (and perhaps in more general hyperK\"ahler
spaces) one can write

\begin{eqnarray}
\label{eq:wLK}
\omega^{\rm LHK}
& = & 
\star_{(4)} dW\, ,
\\
& & \nonumber \\
R^{mn}\wedge R^{nm} 
& = & 
d \omega^{\rm LHK}
=
d \star_{(4)} d W
= 
-\nabla_{(4)}^{2}W  |v| d^{4}x\, ,
\end{eqnarray}

\noindent
where $ |v| d^{4}x$ is the volume 4-form and $W$ is some function defined on
the hyperK\"ahler space and where the subscript $(4)$ indicates that the
operator that carries it is defined in the 4-dimensional hyperK\"ahler
space. For Gibbons-Hawking spaces, which is the only class for which we have
tested this property \cite{Chimento:2018kop}, we get

\begin{equation}
W=(\partial\log{\mathcal{H}})^{2}\, .
\end{equation}

\item Two $\mathrm{SU}(2)$ gauge fields $A^{A_{1,2}}$ defined on the
  hyperK\"ahler manifold whose field strength 2-forms $F^{A_{1,2}}$ are
  self-dual there with respect to the same standard orientation
  (\textit{i.e.}~they are instanton fields)

\begin{equation}
F^{A_{1,2}}= +\star_{(4)}F^{A_{1,2}}\, .  
\end{equation}

\noindent
The most general solution to this equation in an arbitrary hyperK\"ahler space
does not seem to be available in the literature and we will consider some 
particular constructions. All of them, though, have the important
property

\begin{eqnarray}
\label{eq:oYM2}
\omega_{i}^{\rm YM} 
& = &
-\star_{(4)} dV_{i}\, ,   
\\
& & \nonumber \\
F^{A_{i}}\wedge F^{A_{i}} 
& = & 
d \omega_{i}^{\rm YM}
=
-d \star_{(4)} dV_{i}  
= 
\nabla_{(4)}^{2}V_{i} |v| d^{4}x\, ,
\end{eqnarray}

\noindent
for some functions $V_{i}$ defined on the hyperK\"ahler space. These functions
are computed for several instantons in Appendix~\ref{sec:regular_inst}.

\item Three functions $\mathcal{Z}_{0,+,-}$ defined on the hyperK\"ahler space
  which are explicitly given by 

\begin{eqnarray}
\label{eq:Z+}
\mathcal{Z}_{+} 
& = & 
\mathcal{Z}_{+}^{(0)}
-\frac{\alpha'}{2} \left(
\frac{\partial_{m} \mathcal{Z}_{+}^{(0)}\partial_{m}
\mathcal{Z}_{-}^{(0)}}{\mathcal{Z}_{0}^{(0)} \mathcal{Z}_{-}^{(0)}} \right) 
+\mathcal{O}(\alpha'{}^{2})\, ,
\\
& & \nonumber \\
\label{eq:Z-}
\mathcal{Z}_{-} 
& = & 
\mathcal{Z}_{-}^{(0)}
+\mathcal{O}(\alpha'{}^{2})\, ,
\\
& & \nonumber \\
\label{eq:Z0}
\mathcal{Z}_{0} 
& = & 
\mathcal{Z}_{0}^{(0)}
-\frac{\alpha'}{4}
\left[
V_{1}+V_{2}-\left(\partial\log{\mathcal{Z}_{0}^{(0)}}\right)^{2} 
-W \right]
+\mathcal{O}(\alpha'{}^{2})
\, ,
\end{eqnarray}

\noindent
where all the functions with a ${}^{(0)}$ superscript are harmonic in the
hyperK\"ahler space. Notice that the internal product implied in some of the terms
in \eqref{eq:Z+} and \eqref{eq:Z0} is performed using the hyperK\"ahler metric.

\end{enumerate}

Using these building blocks, the remaining fields take the form

\begin{eqnarray}
\label{eq:metric}
ds^{2}
& = &
\frac{2}{\mathcal{Z}_{-}}du
\left[dv-\tfrac{1}{2}\mathcal{Z}_{+} du\right]
-\mathcal{Z}_{0} d\sigma^{2}
-dy^{i}dy^{i}\, ,
\,\,\,\,\,
i,j=1,2,3,4\, ,
\\
& & \nonumber \\
\label{eq:H}
H 
& = & 
d\mathcal{Z}^{-1}_{-}\wedge du \wedge dv
+\star_{(4)}d\mathcal{Z}_{0}\, ,  
\\
& & \nonumber \\
e^{-2{\phi}}
& = &
e^{-2{\phi}_{\infty}}\frac{\mathcal{Z}_{-}}{\mathcal{Z}_{0}}\, .
\end{eqnarray}

\noindent
where $e^{\phi_{\infty}}=g_{s}$. The Kalb-Ramond 2-form $B$ satisfies

\begin{equation}
dB 
= 
d\mathcal{Z}^{-1}_{-}\wedge du \wedge dv
+\star_{(4)}d\mathcal{Z}^{(0)}_{0}\, .
\end{equation}

%%%%%%%%%%%%%%%%%%%%%%%%%%%%%%%%%%%%%%%%%%%%%%%%%%%%%%%%%%%%%%%%%%%%%%
%%%%%%%%%%%%%%%%%%%%%%%%%%%%%%%%%%%%%%%%%%%%%%%%%%%%%%%%%%%%%%%%%%%%%%
%%%%%%%%%%%%%%%%%%%%%%%%%%%%%%%%%%%%%%%%%%%%%%%%%%%%%%%%%%%%%%%%%%%%%%
%%%%%%%%%%%%%%%%%%%%%%%%%%%%%%%%%%%%%%%%%%%%%%%%%%%%%%%%%%%%%%%%%%%%%%
\subsection{The 5-dimensional solutions}
\label{sec-reductionto5dimensions}
%%%%%%%%%%%%%%%%%%%%%%%%%%%%%%%%%%%%%%%%%%%%%%%%%%%%%%%%%%%%%%%%%%%%%%
%%%%%%%%%%%%%%%%%%%%%%%%%%%%%%%%%%%%%%%%%%%%%%%%%%%%%%%%%%%%%%%%%%%%%%
%%%%%%%%%%%%%%%%%%%%%%%%%%%%%%%%%%%%%%%%%%%%%%%%%%%%%%%%%%%%%%%%%%%%%%
%%%%%%%%%%%%%%%%%%%%%%%%%%%%%%%%%%%%%%%%%%%%%%%%%%%%%%%%%%%%%%%%%%%%%%

For our purposes, we only need to know the metric and the two scalar fields of
the 5-dimensional solution (the 5-dimensional dilaton field $\phi$ and the
Kaluza-Klein scalar that measures the radius of the $6\rightarrow 5$
compactification, $k$). These are obtained from the 10-dimensional metric and
dilaton with the same relations used in absence of $\alpha'$ corrections,
\cite{Cano:2017qrq}, and read

\begin{equation}
\label{eq:3chargebh}
\begin{array}{rcl}
ds^{2} 
& = & 
f^{2}dt^{2}-f^{-1}d\sigma^{2}\, ,
\\
& & \\
e^{2\phi}
& = &
e^{2\phi_{\infty}}{\displaystyle\frac{\mathcal{Z}_{0}}{\mathcal{Z}_{-}}}\, ,
\hspace{1cm}
k
= 
k_{\infty}
{\displaystyle\frac{\mathcal{Z}_{+}^{1/2}}{\mathcal{Z}_{0}^{1/4}\mathcal{Z}_{-}^{1/4}}}\, ,
\end{array}
\end{equation}

\noindent 
where $d\sigma^{2}$ is the 4-dimensional hyperK\"ahler metric in
Eq.~(\ref{eq:metricHK}), $\phi_{\infty}$ and $k_{\infty}$ are the asymptotic
values of $\phi$ and $k$, and the metric function $f$ is given by

\begin{equation}
\label{eq:f}
f^{-3}
=
\mathcal{Z}_{0}\, \mathcal{Z}_{+}\, \mathcal{Z}_{-}\, .
\end{equation}

\noindent
The functions $\mathcal{Z}_{0,+,-}$ are given in
Eqs.~(\ref{eq:Z+})-(\ref{eq:Z0}).

We should also mention that the $\mathrm{SU}(2)$ instanton fields have exactly
the same expression as in 10 dimensions.

%%%%%%%%%%%%%%%%%%%%%%%%%%%%%%%%%%%%%%%%%%%%%%%%%%%%%%%%%%%%%%%%%%%%%%
%%%%%%%%%%%%%%%%%%%%%%%%%%%%%%%%%%%%%%%%%%%%%%%%%%%%%%%%%%%%%%%%%%%%%%
%%%%%%%%%%%%%%%%%%%%%%%%%%%%%%%%%%%%%%%%%%%%%%%%%%%%%%%%%%%%%%%%%%%%%%
%%%%%%%%%%%%%%%%%%%%%%%%%%%%%%%%%%%%%%%%%%%%%%%%%%%%%%%%%%%%%%%%%%%%%%
\subsection{The 4-dimensional solutions}
\label{sec-reductionto4dimensions}
%%%%%%%%%%%%%%%%%%%%%%%%%%%%%%%%%%%%%%%%%%%%%%%%%%%%%%%%%%%%%%%%%%%%%%
%%%%%%%%%%%%%%%%%%%%%%%%%%%%%%%%%%%%%%%%%%%%%%%%%%%%%%%%%%%%%%%%%%%%%%
%%%%%%%%%%%%%%%%%%%%%%%%%%%%%%%%%%%%%%%%%%%%%%%%%%%%%%%%%%%%%%%%%%%%%%
%%%%%%%%%%%%%%%%%%%%%%%%%%%%%%%%%%%%%%%%%%%%%%%%%%%%%%%%%%%%%%%%%%%%%%

If the hyperK\"ahler metric $d\sigma^{2}$ in Eq.~(\ref{eq:3chargebh}) is a
Gibbons-Hawking space Eq.~(\ref{eq:GHmetric}) and the other fields of the
5-dimensional solution do not depend on the isometric coordinate $\eta$, we
can dimensionally reduce all fields along that coordinate, obtaining the metric and
scalar fields. However, before doing so, it is convenient to rescale the
coordinate $\eta=R \Psi/2$, where the dimensionless coordinate $\Psi\in
[0,4\pi)$ and the length of the circle is $2\pi R$. The, the Kaluza-Klein
scalar of the $5\to 4$ compactification, that we will denote by $\ell$, has
the asymptotic value $\ell_{\infty}=\tfrac{1}{2}R/\ell_{s}$.  Taking into
account these points we get

\begin{equation}
\label{eq:4chargebh}
\begin{array}{rcl}
ds_{(4)}^{2} 
& = & 
e^{2U}dt^{2}-e^{-2U}d\vec{x}^{\, 2}\, ,
\\
& & \\
e^{2\phi}
& = &
e^{2\phi_{\infty}}{\displaystyle\frac{\mathcal{Z}_{0}}{\mathcal{Z}_{-}}}\, ,
\hspace{1cm}
k
= 
k_{\infty} 
{\displaystyle\frac{\mathcal{Z}_{+}^{1/2}}{\mathcal{Z}_{0}^{1/4}\mathcal{Z}_{-}^{1/4}}}\, ,
\hspace{1cm}
\ell
= 
\ell_{\infty} 
{\displaystyle\frac{\mathcal{Z}_{0}^{1/6}\, \mathcal{Z}_{+}^{1/6}\, \mathcal{Z}_{-}^{1/6}}{\mathcal{H}^{1/2}}}\, , 
\end{array}
\end{equation}

\noindent
where the metric function $e^{-2U}$ is given by 

\begin{equation}
\label{eq:exp2U}
e^{-2U} = \sqrt{\mathcal{Z}_{0}\, \mathcal{Z}_{+}\, \mathcal{Z}_{-} \mathcal{H}}\, .  
\end{equation}

The compactification of the Heterotic Superstring we are considering here
gives an extension of the STU model of $\mathcal{N}=2,d=4$ Supergravity (plus
the $\alpha'$ corrections related to $\Omega_{(-)}$).\footnote{The Yang-Mills
  fields of the Heterotic Superstring appear as non-Abelian vector multiplets
  of the 4-dimensional, gauged, supergravity. } The three scalars above are
the imaginary parts of the three complex scalars of that model.

As for the non-Abelian gauge fields, their reduction follows Kronheimer's
prescription \cite{kn:KronheimerMScThesis}, slightly modified by the
introduction of the parameter $R$. It gives rise to adjoint Higgs fields
$\Phi^{A_{i}}$ and gauge fields $\breve{A}^{A_{i}}$ in $\mathbb{E}^{3}$
related to the components of the gauge field in the Gibbons-Hawking space by

\begin{equation}
\Phi^{A_{i}}
= 
-\mathcal{H} A^{A_{i}}_{\underline{\sharp}}/(R/2)
\hspace{1cm}  
\breve{A}^{A_{i}}_{\underline{x}} 
= 
A^{A_{i}}_{\underline{x}}-\chi_{\underline{x}}
A^{A_{i}}_{\underline{\sharp}}\, .
\end{equation}

The self-duality of the field strength in the hyperK\"ahler space implies the
that $\Phi^{A_{i}}$ and $\breve{A}^{A_{i}}$ are related by the Bogomol'nyi
equation in $\mathbb{E}^{3}$ \cite{Bogomolny:1975de}

\begin{equation}
\label{eq:BogomolnyieqL}
\breve{\mathcal{D}}\Phi^{A_{i}} =\star_{(3)}\breve{F}^{A_{i}}\, .
\end{equation}

\noindent
As shown in \cite{Meessen:2015enl}, the gauge fields constructed in this
way enjoy the 3-dimensional version of the ``Laplacian property'':

\begin{eqnarray}
\omega^{\rm YM}_{i} 
& = &
\star_{(3)}d \left(\Phi^{A_{i}}\Phi^{A_{i}}/\mathcal{H}\right)\, ,  
\\
& & \nonumber \\
F^{A_{i}}\wedge F^{A_{i}} 
& = & 
d\omega^{\rm YM}_{i} 
= 
% -d\star_{(4)}d \left(\Phi^{A_{i}}\Phi^{A_{i}}/H\right)
% = 
\nabla^{2}_{(4)}\left(\Phi^{A_{i}}\Phi^{A_{i}}/\mathcal{H}\right)|v|d^{4}x\, .
\end{eqnarray}

\noindent
This result is reviewed in Appendix~\ref{sec-instanton}.

In the current setup, in order to get a 4-dimensional solution we only need to
choose a set of four harmonic functions
$\mathcal{Z}^{(0)}_{+,-,0},\mathcal{H}$ and a solution of the Bogomol'nyi
equations in $\mathbb{E}^{3}$ (\ref{eq:BogomolnyieqL}). If the solution is, at
zeroth-order in $\alpha'$, a single (spherically-symmetric),
asymptotically-flat, regular, extremal black hole, the functions
$\mathcal{Z}^{(0)}_{+,-,0}$ are of the form

\begin{equation}
\label{eq:harmonic_Z}
\mathcal{Z}^{(0)}_{+,-,0} 
=
1 +\frac{q_{+,-,0}}{r}\, ,  
\end{equation}

\noindent
and we are bound to choose\footnote{Multicenter solutions demand multicenter
  $\mathcal{Z}^{(0)}_{+,-,0}$s and multicenter $\mathcal{H}$s, but this case
  will be studied elsewhere.}

\begin{equation}
\label{eq:HforKKmonopole}
\mathcal{H}= 1 +\frac{q}{r}\, ,
\end{equation}

\noindent
where $r=|\vec{x}|$ and $q$ plays the r\^ole of a $\mathrm{U}(1)$ magnetic
charge in $d=4$.  The corresponding Gibbons-Hawking space is a Kaluza-Klein
(KK) monopole (also known as Euclidean Taub-NUT space) which, for large values
of $r$, asymptotes to $\mathbb{E}^{3}\times \mathrm{S}^{1}$. For this reason,
the KK monopole is not used to construct asymptotically-flat 5-dimensional
solutions and $\mathcal{H}=1/r$, which corresponds to
$\mathbb{E}^{4}_{-\{0\}}$, is used instead in that case. Since the $q/r$ term
dominates in the $r\to 0$ limit, the 4-dimensional solutions have a
5-dimensional core even though asymptotically the have only four non-compact
dimensions.

As for the solutions of Eq.~(\ref{eq:BogomolnyieqL}) (by definition, BPS
magnetic monopoles, among which the 't~Hooft-Polyakov magnetic monopole
\cite{tHooft:1974kcl,Polyakov:1974ek} in the Prasad-Sommerfield limit
\cite{Prasad:1975kr} is a particular example), we must look for
spherically-symmetric field configurations. Fortunately, all of them where
found by Protogenov in Ref.~\cite{Protogenov:1977tq}.  Independently of their
singular character in $\mathbb{E}^{3}$, all of them have been used to
construct regular, extremal, spherically-symmetric black holes in $d=4$
dimensions and the 't~Hooft-Polyakov BPS monopole has been used to construct
globally regular solutions
\cite{Huebscher:2007hj,Meessen:2008kb,Hubscher:2008yz,Bueno:2014mea,Meessen:2015nla}.\footnote{$\mathbb{E}^{3}$
  is just an auxiliary space. The complete physical solution can be regular
  even if one uses a monopole solution which is singular in that auxiliary
  space. Typically, the singularity of the monopole in $\mathbb{E}^{3}$ is
  resolved by the extremal horizon.}

Using these BPS monopole solutions and Kronheimer's prescription we get
selfdual instanton fields in 5 and 10 dimensions and we want them to be regular.
For a given BPS monopole, this property depends critically on the choice of
$\mathcal{H}$ and, more precisely, on its singularities and the behavior of
the adjoint Higgs field $\Phi^{A}$ at those singularities. In
Ref.~\cite{kn:KronheimerMScThesis} Kronheimer gave the conditions under which
the contributions to the instanton number (action) in a neighborhood of the
singularities of $\mathcal{H}$ are finite. It is easy to see that all the monopoles
found by Protogenov satisfy them if $\mathcal{H}$ is also
spherically-symmetric, that is, if $\mathcal{H}=a+b/r$ for two positive
constants $ab\neq 0$.  

Observe that, even if Kronheimer's conditions are satisfied at $r=0$, the
instanton number density $F^{A_{i}}\wedge F^{A_{i}}$ may not fall off fast
enough for large values of $r$ to give a finite integral. For
$\mathcal{H}=1/r$, the choice that leads to asymptotically-flat 5-dimensional
solutions, this problem was shown in Ref.~\cite{Bueno:2015wva} to arise for
all of Protogenov's solutions except for the 1-parameter family of
``colored'' monopoles discussed in
Refs.~\cite{Meessen:2008kb,Meessen:2015nla}, which turn out to correspond to
the BPST instanton \cite{Belavin:1975fg}. These instantons were used in
Ref.~\cite{Meessen:2015enl} as part of the solution-generating technique found
in Ref.~\cite{Bellorin:2007yp} to construct black holes with non-Abelian
hair and, in Ref.~\cite{Cano:2017sqy} to construct a globally regular
instanton solution which corresponds to the compactification of the
10-dimensional heterotic ``gauge 5-brane'' of Ref.~\cite{Strominger:1990et}.

For $\mathcal{H}=1+q/r$, all the monopoles found by Protogenov give instantons
over the KK monopole with good asymptotic behavior. We present them in detail
in the next section. Then, in the following sections we study the 4- 5- and
10-dimensional solutions they give rise to.

%%%%%%%%%%%%%%%%%%%%%%%%%%%%%%%%%%%%%%%%%%%%%%%%%%%%%%%%%%%%%%%%%%%%%%
%%%%%%%%%%%%%%%%%%%%%%%%%%%%%%%%%%%%%%%%%%%%%%%%%%%%%%%%%%%%%%%%%%%%%%
%%%%%%%%%%%%%%%%%%%%%%%%%%%%%%%%%%%%%%%%%%%%%%%%%%%%%%%%%%%%%%%%%%%%%%
%%%%%%%%%%%%%%%%%%%%%%%%%%%%%%%%%%%%%%%%%%%%%%%%%%%%%%%%%%%%%%%%%%%%%%
\section{Ingredients of the 4-dimensional black hole solution}
\label{Ingredients}
%%%%%%%%%%%%%%%%%%%%%%%%%%%%%%%%%%%%%%%%%%%%%%%%%%%%%%%%%%%%%%%%%%%%%%
%%%%%%%%%%%%%%%%%%%%%%%%%%%%%%%%%%%%%%%%%%%%%%%%%%%%%%%%%%%%%%%%%%%%%%
%%%%%%%%%%%%%%%%%%%%%%%%%%%%%%%%%%%%%%%%%%%%%%%%%%%%%%%%%%%%%%%%%%%%%%
%%%%%%%%%%%%%%%%%%%%%%%%%%%%%%%%%%%%%%%%%%%%%%%%%%%%%%%%%%%%%%%%%%%%%%

%%%%%%%%%%%%%%%%%%%%%%%%%%%%%%%%%%%%%%%%%%%%%%%%%%%%%%%%%%%%%%%%%%%%%%
%%%%%%%%%%%%%%%%%%%%%%%%%%%%%%%%%%%%%%%%%%%%%%%%%%%%%%%%%%%%%%%%%%%%%%
%%%%%%%%%%%%%%%%%%%%%%%%%%%%%%%%%%%%%%%%%%%%%%%%%%%%%%%%%%%%%%%%%%%%%%
%%%%%%%%%%%%%%%%%%%%%%%%%%%%%%%%%%%%%%%%%%%%%%%%%%%%%%%%%%%%%%%%%%%%%%
\subsection{KK monopole of arbitrary charge}
%%%%%%%%%%%%%%%%%%%%%%%%%%%%%%%%%%%%%%%%%%%%%%%%%%%%%%%%%%%%%%%%%%%%%%
%%%%%%%%%%%%%%%%%%%%%%%%%%%%%%%%%%%%%%%%%%%%%%%%%%%%%%%%%%%%%%%%%%%%%%
%%%%%%%%%%%%%%%%%%%%%%%%%%%%%%%%%%%%%%%%%%%%%%%%%%%%%%%%%%%%%%%%%%%%%%
%%%%%%%%%%%%%%%%%%%%%%%%%%%%%%%%%%%%%%%%%%%%%%%%%%%%%%%%%%%%%%%%%%%%%%

We consider the Gibbons-Hawking metric 

\begin{equation}\label{GH1}
d\sigma^{2} =  
\mathcal{H}^{-1}(d\eta+\chi)^{2}
+\mathcal{H}dx^{i}dx^{i}\, ,\quad \text{where}\,\, d\mathcal{H}=\star_{(3)}d\chi\, .
\end{equation}

\noindent
Here $\eta$ is a compact coordinate of periodicity $\eta\sim \eta+2\pi R$. We
choose the following harmonic function $\mathcal{H}$

\begin{equation}
\mathcal{H}=1+\frac{q}{r}\, .
\end{equation}

It is convenient to use spherical coordinates $\theta,\phi$ defined by

\begin{equation}
\frac{x^{1}}{r}=\sin{\theta}\cos{\phi}\, ,
\quad 
\frac{x^{2}}{r}=-\sin{\theta}\sin{\phi}\, ,
\quad 
\frac{x^{3}}{r}=-\cos{\theta}\, ,
\end{equation}

\noindent
so that, locally, the 1-form $\chi$ reads

\begin{equation}\label{chi-1}
\chi=q \cos{\theta} d\phi\, . 
\end{equation}

\noindent
In these coordinates, the metric of the KK monopole takes the
form

\begin{eqnarray}
d\sigma^{2}
& = & 
\mathcal{H}^{-1}(d\eta+q\cos{\theta} d\phi)^{2}
+\mathcal{H}\left(dr^{2}+r^{2}d\Omega^{2}_{(2)}\right)\, ,
\end{eqnarray}

\noindent
where 

\begin{equation}
d\Omega^{2}_{(2)}
=
d\theta^{2}+\sin^{2}\theta d\phi^{2}\, ,  
\end{equation}

\noindent
is the metric of the round $\mathrm{S}^{2}$ of unit radius. However, a global
description of the solution requires two patches.  The 1-form
$\chi=q\cos{\theta} d\phi$ contains a Dirac-Misner string at the poles
$\theta=0,\pi$, or equivalently, in the line $x^{1}=x^{2}=0$. This can be easily
checked by computing the norm of $\chi$:

\begin{equation}
|\cos{\theta} d\phi|^{2}=q^{2}\frac{\cot^{2}{\theta}}{\mathcal{H} r^{2}}\, ,
\end{equation}

\noindent
which is divergent at those points. In order to fix this singularity, we work with 
two different patches:

\begin{equation}
\chi^{(+)}=q\left(\cos{\theta}-1\right)d\phi\, , 
\quad 
\chi^{(-)}=q\left(\cos{\theta}+1\right)d\phi\, .
\end{equation}

\noindent
In this way, $\chi^{(+)}$ is regular everywhere except at $\theta=\pi$, and
$\chi^{(-)}$ is regular everywhere except at $\theta=0$. We also have to use
different coordinates $\eta^{(+)}$ and $\eta^{(-)}$ in every patch, but in the
intersection we must have

\begin{equation}
d\eta^{(+)}+\chi^{(+)}=d\eta^{(-)}+\chi^{(-)}
\Rightarrow
d\left(\eta^{(+)}-\eta^{(-)}\right)=2qd\phi\, .
\end{equation}

\noindent
Hence, we conclude that 

\begin{equation}
\eta^{(+)}-\eta^{(-)}=2q\phi\, .
\end{equation}

\noindent
Since $\phi$ has period $2\pi$ and both $\eta^{(\pm)}$ have period $2\pi R$ by
definition, this relation can only hold if $q$ satisfies the quantization
condition

\begin{equation}
\label{quant}
q=\frac{n R}{2}\, , \quad n=1,2,\ldots
\end{equation}

\noindent
The reason is that $\eta\sim \eta+2\pi R$ trivially implies that $\eta\sim
\eta+2\pi n R$, and thus the Dirac-Misner string is avoided for all
$n=1,2\ldots$ Let us then introduce the angular coordinate

\begin{equation}
\Psi=\frac{2\eta}{R}\Rightarrow \Psi \sim\Psi+4\pi\, .
\end{equation}

\noindent
Taking into account the quantization of the charge $q$, we can write then the
metric (locally) as

\begin{eqnarray}
d\sigma^{2}
& = & 
\mathcal{H}^{-1}\frac{R^{2}}{4}(d\Psi+n \cos{\theta} d\phi)^{2}
+\mathcal{H}\left(dr^{2}+r^{2}d\Omega^{2}_{(2)}\right)\, .
\end{eqnarray}

It is important for future purposes to understand the $r\rightarrow 0$ and
$r\rightarrow \infty$ limits of this space. 

\begin{itemize}
\item In the $r\rightarrow 0$ limit we must use that $\mathcal{H}\sim\frac{n
    R}{2r}$, and after performing the change of coordinates

  \begin{equation}
    \label{r-rho}
    r=\frac{\rho^{2}}{2nR}\, ,
  \end{equation}

\noindent
we obtain

\begin{equation}
  d\sigma^{2}(r\rightarrow0)
  \sim
  d\rho^{2}+\frac{\rho^{2}}{4}\left[\left(\frac{d\Psi}{n}
      +\cos{\theta} d\phi\right)^{2}
    +d\theta^{2}+\sin^{2}\theta d\phi^{2}\right]\, .
\end{equation}

When $n=1$, we recognize the factor that $\rho^{2}$ multiplies as the metric
of the round $\mathrm{S}^{3}$.  However, for $n>1$ the cyclic coordinate
$\Psi$ does not cover the full sphere, but only a $1/n$ part of it. This
corresponds to the metric of a lens space $\mathrm{S}^{3}/\mathbb{Z}_{n}$, and
hence the full space near $r=0$ is the orbifold
$\mathbb{E}^{4}/\mathbb{Z}_{n}$. Although lens spaces are regular, the full
Gibbons-Hawking metric contains a conical singularity at $r=0$, because at
this point the periodicity of $\Psi$ is not ``the right one'' for
$n>1$. Nevertheless, it is important to notice that, when one takes into
account the full 10-dimensional metric with non-vanishing $q_{0}$ there is no
conical singularity, see Eqs.~(\ref{eq:metric}) and (\ref{eq:harmonic_Z}).
 
\item In the asymptotic limit $r\rightarrow \infty$ we have
  $\mathcal{H}\rightarrow 1$ and the metric becomes the direct product
  $\mathrm{S}^{1}\times\mathbb{E}^{3}$:

\begin{equation}
  \label{GHinf}
  d\sigma^{2}(r\rightarrow\infty)=d\eta^{2}+dx^{i}dx^{i}\, .
\end{equation}

This is better seen by using Cartesian coordinates $x^{i}$ and the two patches
introduced previously.  In that case, the 1-forms $\chi^{(\pm)}$ read

\begin{equation}
  \chi^{(\pm)}=q\frac{x^{1}dx^{2}-x^{2} dx^{1}}{r(x^{3}\pm r)}\, ,
  \quad 
  \text{where}\,\,\, r=\sqrt{(x^{1})^{2}+(x^{2})^{2}+(x^{3})^{2}}\, .
\end{equation}

\noindent
We use $\chi^{(+)}$ in the upper space $x^{3}\ge 0$ and $\chi^{(-)}$ in the
lower one $x^{3}\le 0$. In this way, it is explicit that $\chi^{(\pm)}$ are
regular in their respective regions, and we observe that
$\lim_{r\rightarrow\infty}\chi^{(\pm)}=0$, where the limit is again taken in
the respective region.  Hence, the metric (\ref{GH1}) becomes (\ref{GHinf}).

\end{itemize}

%%%%%%%%%%%%%%%%%%%%%%%%%%%%%%%%%%%%%%%%%%%%%%%%%%%%%%%%%%%%%%%%%%%%%%
%%%%%%%%%%%%%%%%%%%%%%%%%%%%%%%%%%%%%%%%%%%%%%%%%%%%%%%%%%%%%%%%%%%%%%
%%%%%%%%%%%%%%%%%%%%%%%%%%%%%%%%%%%%%%%%%%%%%%%%%%%%%%%%%%%%%%%%%%%%%%
%%%%%%%%%%%%%%%%%%%%%%%%%%%%%%%%%%%%%%%%%%%%%%%%%%%%%%%%%%%%%%%%%%%%%%
\subsection{$\mathrm{SU}(2)$ instantons over KK monopoles}
\label{Instantons}
%%%%%%%%%%%%%%%%%%%%%%%%%%%%%%%%%%%%%%%%%%%%%%%%%%%%%%%%%%%%%%%%%%%%%%
%%%%%%%%%%%%%%%%%%%%%%%%%%%%%%%%%%%%%%%%%%%%%%%%%%%%%%%%%%%%%%%%%%%%%%
%%%%%%%%%%%%%%%%%%%%%%%%%%%%%%%%%%%%%%%%%%%%%%%%%%%%%%%%%%%%%%%%%%%%%%
%%%%%%%%%%%%%%%%%%%%%%%%%%%%%%%%%%%%%%%%%%%%%%%%%%%%%%%%%%%%%%%%%%%%%%

The BPS monopoles and the corresponding instantons that we consider can be
written in terms of the functions $f(r),h(r)$ by

\begin{align}
\Phi^{A} 
& = 
-\frac{x^{A}}{r} (r f)\, ,
\hspace{1cm}
\breve{A}^{A}{}_{\underline{x}} 
= 
-\epsilon^{A}{}_{xz} \frac{x^{z}}{r} (rh)\, ,    
\\
& \nonumber \\ \label{vectorAA}
A^{A}
 &=
\mathcal{H}^{-1}\frac{x^{A}}{r}\,  (rf)\, \frac{R}{2}(d\Psi+n \cos\theta d\phi)
-\epsilon^{A}{}_{xz}r^{2}
h\frac{x^{z}}{r}d\left(\frac{x^{x}}{r}\right)\ .
\end{align}

There are two independent families of solutions. One of them corresponds to the 
\textit{colored monopole}, which depends on a parameter $\lambda$:

\begin{equation}
f_{\lambda}(r)
=
h_{\lambda}(r)
=
-\frac{1}{gr^{2}}\frac{1}{1+\lambda^{2} r}\, .
\end{equation} 

\noindent
The other family depends on two parameters $(\mu, s)$, and it is given by

\begin{equation}
rf_{\mu,s}(r)
=
-\frac{1}{gr}\left[1-\mu r \coth(\mu r+s)\right]\, ,
\quad  
rh_{\mu,s}(r)
=
-\frac{1}{gr}\left[1-\frac{\mu r}{\sinh(\mu r+s)}\right]\, .
\end{equation} 

The parameter $\mu$ can always be taken to be positive, while $s$, which in
this context is known as \textit{Protogenov hair parameter}, can take any real
value. However, we will only consider $s\geq 0$ to avoid singularities.

The $s=0$ member of this family is the 't~Hooft-Polyakov magnetic monopole
\cite{tHooft:1974kcl,Polyakov:1974ek} in the BPS limit \cite{Prasad:1975kr}
with mass parameter $\mu$ and it is the only regular monopole in this family.
On the other hand, in the  $s\rightarrow \infty$ limit we get

\begin{equation}
rf_{\mu,\infty}(r)
=
\frac{\mu}{g}-\frac{1}{gr} \, ,
\quad  
rh_{\mu,\infty}(r)= -\frac{1}{gr}\, ,
\end{equation} 

\noindent
which is a $\mu$-dependent generalization of the Wu-Yang $\mathrm{SU}(2)$
monopole \cite{Wu:1967vp}.

We are going to characterize the instantons over a KK monopole that all these
monopoles give rise to. Observe, first of all, that the norm of the Higgs
field is given by

\begin{equation}
|\Phi|=(\Phi^{A}\Phi^{A})^{1/2} = |rf|\, ,
\end{equation}

\noindent
and that Kronheimer's conditions are satisfied:

\begin{equation}
\lim_{r\to 0 }r|\Phi_{\lambda}| 
= \frac{1}{g}\, ,
\hspace{1cm}
\lim_{r\to 0 } d(r|\Phi_{\lambda}|) 
= -\frac{\lambda^{2}}{g}dr
\end{equation}

\noindent
for the colored monopole, and

\begin{equation}
\lim_{r\to 0 }r|\Phi_{\mu,s}| 
= 
\left\{
  \begin{array}{lr}
{\displaystyle\frac{1}{g}}\, ,\,\,\,\, & s\neq 0\, , \\
\\
0\, ,\,\,\,\,& s=0\, ,\\
\end{array}
\right.   
\hspace{1cm}
\lim_{r\to 0 } d(r|\Phi_{\mu,s}|) 
= 
\left\{
  \begin{array}{cr}
{\displaystyle -\frac{\mu\coth{s}}{g}}dr\, ,\,\,\,\, &  s\neq 0\, , \\
\\
0\, ,\,\,\,\, & s=0\, \\
\end{array}
\right.   
\end{equation}

\noindent
for the $(\mu,s)$ family.\\

Except for the $(\mu,s=0)$ case, the vector fields $A^{A}$ in
Eq.~(\ref{vectorAA}) are singular at $r=0$ in the Gibbons-Hawking space
(\ref{GH1}). The fact that Kronheimer's conditions are satisfied ensures that
these singularities can be removed by performing an appropriate gauge
transformation (in fact, the field strength $F^{A}$ is regular). It is not too
difficult to show that this is the case when $n=1$, for all the instantons
constructed from the above monopoles. However, for $n>1$, the presence of a
conical singularity in the GH space at $r=0$ makes the problem more
complicated, and we have not found the gauge in which the vectors become
regular when defined over that space. Nevertheless, we stress that the
physical fields are not defined over the GH space, but over the physical
manifold whose metric is given by Eq.~(\ref{eq:metric}).  As explained in the
previous section, in the full spacetime the conical singularity at $r=0$ is
blown away by the conformal factor in front of the hyperK\"ahler metric. This
mechanism also resolves the original divergences of these vectors, rendering
the instanton fields regular.

The regularity conditions on the $\mathrm{SU}(2)$ gauge fields have relevant
physical implications. For example, in the $(\mu, s)$ family, the parameter
$\mu$ is related to the charge of the instanton. However, from the String
Theory point of view, the charge cannot depend on an arbitrary continuous
parameter.  As we explain in Appendix~\ref{sec:regular_inst}, the correct
construction of this family of solutions implies that the parameter $\mu$
should actually be quantized \cite{BoutalebJoutei:1979iy}, according to
Eq.~(\ref{mmun}), which we write here for convenience

\begin{equation}
\mu=\frac{2m}{R n}, \quad m=0,1,2,\ldots 
\end{equation}

In the case of the $\lambda$-family of instanton solutions built from the
colored monopoles, the charge is independent of any continuous parameter, as 
we show below.  Moreover, as we will see in the following section, the
instantons of the $\lambda$-family are the most interesting ones from the
point of view of $\alpha'$ corrections.

The ``charge'' of these instantons is just their instanton number, which is given
by

\begin{equation}
\mathfrak{n}
=
-\frac{g^{2}}{8\pi^{2}}\int 
\mathrm{Tr}\left[\hat{F}\wedge \hat{F}\right]
=
\frac{g^{2}}{16\pi^{2}}\int F^{A}\wedge F^{A}\, ,
\end{equation}

\noindent
where the integral is taken over the GH space.\footnote{More precisely, it is
  defined over the 4-dimensional space conformal to the GH space, but the
  integral is invariant under Weyl transformations.} This integral takes a
simple form once we take into account that $F^{A}\wedge F^{A}$ is the
Laplacian of some function $V$ (see Appendix~\ref{sec-instanton}):

\begin{equation}
F^{A}\wedge F^{A}=-d\star_{(4)}d V\, .
\end{equation}

\noindent
Using that relation, for the cases that we consider $V=V(r)$, we can write

\begin{equation}
d\star_{(4)}d V=\star_{(4)}\frac{1}{|v|}\partial_{r}\left(\mathcal{H}^{-1}|v|\partial_{r} V\right)\, ,
\end{equation}

\noindent
and integration yields

\begin{equation}
\mathfrak{n}
=
-\frac{g^{2}}{16\pi^{2}}
\int 
\frac{R}{2} d\Psi\wedge d\theta\wedge d\phi
\sin\theta r^{2}\partial_{r} V\Big|^{r=0}_{r=\infty}
=
-\frac{g^{2}R}{2}r^{2}\partial_{r} V\Big|^{r=0}_{r=\infty} \, .
\end{equation}

\noindent 
For our instantons the function $V$ is given by (see
Appendix~\ref{sec-instanton})

\begin{equation}
V
=
\frac{\Phi^{A}\Phi^{A}}{\mathcal{H}}=\frac{r^{2} f(r)^{2}}{\mathcal{H}}\, ,
\end{equation}

\noindent
and we obtain

\begin{equation}
\label{eq:instnumber}
\mathfrak{n}_{\lambda}
=
\frac{1}{n}\, ,
\qquad 
\mathfrak{n}_{\mu,s}
=
\frac{\left(m+1\right)^{2}-\delta_{s,0}}{n}\, .
\end{equation}

As we see, the instanton number is quantized although its value is not
necessarily an integer. This is related to the presence of the lens space
$\mathrm{S}^{3}/\mathbb{Z}_{n}$ and it shows that these instantons, when
$n>1$, are somewhat exotic in a mathematical sense. Other solutions with
rational but discrete instanton number are known in the literature, see, for
instance, Refs.~\cite{BoutalebJoutei:1979iy, BoutalebJoutei:1979iz}.

%%%%%%%%%%%%%%%%%%%%%%%%%%%%%%%%%%%%%%%%%%%%%%%%%%%%%%%%%%%%%%%%%%%%%%
%%%%%%%%%%%%%%%%%%%%%%%%%%%%%%%%%%%%%%%%%%%%%%%%%%%%%%%%%%%%%%%%%%%%%%
%%%%%%%%%%%%%%%%%%%%%%%%%%%%%%%%%%%%%%%%%%%%%%%%%%%%%%%%%%%%%%%%%%%%%%
%%%%%%%%%%%%%%%%%%%%%%%%%%%%%%%%%%%%%%%%%%%%%%%%%%%%%%%%%%%%%%%%%%%%%%
\section{Explicit 4-dimensional black hole solutions and their charges}
\label{sec-all}
%%%%%%%%%%%%%%%%%%%%%%%%%%%%%%%%%%%%%%%%%%%%%%%%%%%%%%%%%%%%%%%%%%%%%%
%%%%%%%%%%%%%%%%%%%%%%%%%%%%%%%%%%%%%%%%%%%%%%%%%%%%%%%%%%%%%%%%%%%%%%
%%%%%%%%%%%%%%%%%%%%%%%%%%%%%%%%%%%%%%%%%%%%%%%%%%%%%%%%%%%%%%%%%%%%%%
%%%%%%%%%%%%%%%%%%%%%%%%%%%%%%%%%%%%%%%%%%%%%%%%%%%%%%%%%%%%%%%%%%%%%%
 
Having described all the basic building blocks necessary to construct a
solution, we just need to specify our choices for them. 

We choose single-pole, spherically symmetric, harmonic functions
$\mathcal{Z}_{0,+,-}$ of the form Eqs.~(\ref{eq:harmonic_Z}) and the GH space
of the charge-$q$ KK monopole described in the last section with harmonic
function $\mathcal{H}$ given by Eq.~(\ref{eq:HforKKmonopole}). In addition, we
are going to include an arbitrary number $N_{\lambda}$ and $N_{\mu,s}$ of
instantons of the $\lambda$ and $(\mu, s)$ families, respectively, which we
have described in Section~\ref{Instantons}.  Introducing this input in
Eqs.~(\ref{eq:Z+})-(\ref{eq:Z0}) and using the relations
Eqs.~(\ref{eq:instZ0a})-(\ref{eq:instZ0c}) for the contribution from the
instantons $V_{1},V_{2}$, we obtain the $\alpha'$-corrected functions

\begin{align}
\label{eq:exp_Z+}
\mathcal{Z}_{+} 
& = 
1+\frac{q_{+}}{r}- \frac{\alpha'}{2} \frac{q_{+} q_{-} }{r (r+q)(r+q_{0})(r+q_{-})}
 +\mathcal{O}(\alpha'{}^{2})\, ,
\\
& \nonumber\\
\label{eq:exp_Z-}
\mathcal{Z}_{-} 
& = 
1+\frac{q_{-}}{r}+\mathcal{O}(\alpha'{}^{2})\, ,
\\
& \nonumber\\
\label{eq:exp_Z0}
\mathcal{Z}_{0} 
& = 
1+\frac{q_{0}}{r}
\\ & 
+ \frac{\alpha'}{4r(r+q)}
\Bigg\{\frac{q_{0}^{2}}{(r+q_{0})^{2}}
+\frac{q^{2}}{(r+q)^{2}}-\sum_{i=1}^{N_{\lambda}}\frac{1}{(1+\lambda_{i}^{2}r)^{2}}
-\sum_{i=1}^{N_{\mu,s}}\left[ 1-\mu_i r \coth \left(\mu_i r + s_i \right) \right]^2
\Bigg\} \, ,
\\
\label{eq:exp_H}
\mathcal{H}
& = 
1+\frac{q}{r}+\mathcal{O}(\alpha'{}^{2})\, ,
\end{align}

\noindent
where we recall that the charge $q$ is quantized according to
Eq.~(\ref{quant}). Note, however, that these functions are not univocally
determined: we are free to add an arbitrary $\mathcal{O}(\alpha')$ harmonic
function to each of them, and the resulting field configuration is still a
solution of the equations of motion at first order in $\alpha'$. We use this
freedom to impose that the $1/r$ pole of the $\mathcal{Z}$ functions is not
changed by the $\alpha'$ corrections and to ensure that
$\mathcal{Z}\rightarrow 1$ at infinity. For the functions above this amounts
to the changes

\begin{equation}
\label{eq:subtractedZs}
\mathcal{Z}_{+}
\to
\mathcal{Z}_{+} + \frac{\alpha'}{2r q}\frac{q_{+}}{q_{0}}\,,
\hspace{2cm}
\mathcal{Z}_{0}
\to
\mathcal{Z}_{0}- \frac{\alpha'}{4r q}
\left[2-N_\lambda-N_{\mu,s}+\sum_{i=1}^{N_{\mu,s}}
\left(\delta_{s,0} +rq\mu_{i}^{2} \right) \right]\, .
\end{equation}

\noindent
There are two reasons why we must eliminate the poles from the $\alpha'$
corrections:

\begin{enumerate}
\item The $\alpha'$ corrections are associated to the curvatures of the gauge
  instantons and torsionful spin connection, which are regular. Thus, they
  should be regular as well. The poles are spurious and their presence is
  solely due to the fact that we are using a singular gauge to write the
  different connections.
\item We want to associate the residues of the poles with the sources of the
  solution, and these should not be modified by the $\alpha'$ corrections.
\end{enumerate}

From now on we focus the discussion on the gauge fields on the
$\lambda$-instantons, which will shortly be proven as the most interesting
family. Then, for simplicity reasons, we will set $N_{\mu,s}=0$. Taking into
account all these points, the functions that determine the solution read

\begin{align}
\label{eq:exp_Z+2}
\mathcal{Z}_{+} 
& = 
1+\frac{q_{+}}{r}+ \frac{\alpha' q_{+}}{2qq_{0}} 
\frac{r^{2}+r(q_{0}+q_{-}+q)+q q_{0}+q q_{-}+q_{0}q_{-}
}{(r+q)(r+q_{0})(r+q_{-})} 
+\mathcal{O}(\alpha'{}^{2})\, ,
\\
& \nonumber\\
\label{eq:exp_Z-2}
\mathcal{Z}_{-} 
& = 
1+\frac{q_{-}}{r}+\mathcal{O}(\alpha'{}^{2})\, ,
\\
& \nonumber\\
\label{eq:exp_Z02}
\mathcal{Z}_{0} 
& = 
1+\frac{q_{0}}{r}+ \alpha'\Bigg\{-F(r;q_{0})-F(r;q)+\sum_{i=1}^{N_{\lambda}}F
(r;\lambda_{i}^{-2})
%+\sum_{i=1}^{N_{\mu,s}}G(r;\mu_{i},s_{i})
\Bigg\} +\mathcal{O}(\alpha'{}^{2}) \, ,
\\
& \nonumber\\
\label{eq:exp_H2}
\mathcal{H}
& =
1+\frac{q}{r}+\mathcal{O}(\alpha'{}^{2})\, ,
\end{align}

\noindent
where we have introduced the function

\begin{align}
F(r;k) 
&:=\frac{(r+q)(r+2k)+k^{2}}{4q(r+q)(r+k)^{2}}\, ,
%\\
%G(r;\mu_{i},s_{i}) 
%& :=\mu_{i}^{2}+\frac{1-\delta_{s_{i},0}}{r q}
%-\frac{\left[ 1-\mu_{i} r coth \left(\mu_{i} r + s_{i} \right) \right]^{2}}{r(r+q)} \, .
\end{align}

Expressed in this way, it is obvious that we can eliminate all the $\alpha'$
corrections to $\mathcal{Z}_{0}$ if we use $N_{\lambda}=2$ instantons of sizes
$\lambda_{1}^{-2}=q_{0}$, $\lambda_{2}^{-2}=q$ ($N_{\mu,s}=0$). We will come
back to this point later.

In the configuration at hand, $q_{0}$ is related to the number of solitonic
(or Neveu-Schwarz, NS) 5-branes (S5), $q_{-}$ is related to the winding number
of a string wrapped along the $u$ direction (F1) and $q_{+}$ represents the
momentum of a wave (W) along that direction. We also have a KK monopole of
charge $n$ ($q=n R/2$) and a number $N_{\lambda}$ of gauge 5-branes, sourced by the
$\mathrm{SU}(2)$ instantons.

The easiest way to determine the number of stringy objects is by looking at
the near-horizon geometry $r\rightarrow 0$. In that case, we introduce the
coordinate $\rho$ in (\ref{r-rho}) such that $d\sigma^{2}$ becomes explicitly
$\mathbb{E}^{4}/\mathbb{Z}_{n}$. The full 10-dimensional geometry is regular
and corresponds to the spacetime geometry $\mathrm{AdS}_{3}\times
\mathrm{S}^{3}/\mathbb{Z}_{n}\times \mathrm{T}^{4}$. The $\mathcal{Z}$
functions behave in that limit as

\begin{equation}
\mathcal{Z}_{0}
\sim
\frac{2nR q_{0}}{\rho^{2}}\, ,
\quad 
\mathcal{Z}_{-}\sim \frac{2nR q_{-}}{\rho^{2}}\, ,
\quad 
\mathcal{Z}_{+}\sim \frac{2nR q_{+}}{\rho^{2}}\, .
\end{equation}

\noindent
This near-horizon geometry is the same as the one of the 5-dimensional black
holes considered in Refs.~\cite{Cano:2017qrq,Cano:2018qev} up to a
$\mathbb{Z}_{n}$ quotient, and therefore we can apply the results there in
order to obtain the number of stringy objects from the coefficients of
$1/\rho^{2}$. There is one difference though: here we only have $1/n$-th of
the sphere, and this means that the field produced by one of these objects is
$n$ times larger than in the case in which we have the full sphere. Taking
into account this effect, we expect the result to be

\begin{equation}\label{eq:charges}
q_{0}
=
\frac{\ell_{s}^{2}}{2R}N_{S5}\, ,
\qquad 
q_{-}=\frac{\ell_{s}^{2}g_{s}^{2}}{2R}N_{F1}\, ,
\qquad 
q_{+}=\frac{g_{s}^{2}\ell_{s}^{4}}{2R R_{u}^{2}}N_{W}\, .
\end{equation}

Indeed, one can check that the values of $q_{0}$ and $q_{-}$ above agree with
the ones computed by using the relations\footnote{Here the tensions read
\begin{equation}
  T_{S5}=\frac{1}{(2\pi \ell_{s})^{5}\ell_{s}g_{s}^{2}}\, ,
\hspace{1cm}
  T_{F1}=\frac{1}{2\pi\alpha'}\, , 
\end{equation}
and the NSNS 7-form field strength $\tilde{H}$ is defined as $ \tilde{H}=\star
e^{-2\phi} H$.}

\begin{eqnarray}
g_{s}^{2} N_{S5}T_{S5}
& = &
\frac{g_{s}^{2}}{16\pi G_{N}^{(10)}} \left\{\int_{\mathrm{S}\times\mathrm{S}_{\infty}^{2}}\star e^{2\phi}\tilde{H}
-\frac{\alpha'}{4}\int\left( F^{A}\wedge F^{A}+{R}_{(-)}{}^{{a}}{}_{{b}}\wedge
{R}_{(-)}{}^{{b}}{}_{{a}}\right)\right\}\, ,\,\,\,\,\,\,\,
\\
& & \nonumber \\
T_{F1}N_{F1}
& = & 
\frac{g_{s}^{2}}{16\pi G_{N}^{(10)}}
\int_{\mathrm{S}\times\mathrm{S}_{\infty}^{2}\times \mathrm{T}^{4}}\star e^{-2\phi}H\, , 
\end{eqnarray}

\noindent
obtained by coupling the 10-dimensional Heterotic Superstring effective action
to the worldvolume effective actions of $N_{S5}$ solitonic 5-branes and
$N_{F1}$ fundamental strings~\cite{Cano:2017qrq}.

The interpretation of $q_{+}$ is, however, less transparent. This is related
to the fact that the parameters $q_{0}$, $q_{-}$ and $q_{+}$ represent
localized sources of solitonic 5-brane, string and momentum charge
respectively. But due to the effect of the $\alpha'$ corrections, those do not
necessarily amount for the corresponding total charges measured at
infinity. For instance, when $r\rightarrow \infty$, the functions
$\mathcal{Z}_{+,0}$ take the form

\begin{align}
\mathcal{Z}_{+} 
& = 
1+\frac{1}{r}\left(q_{+}+\frac{\alpha' q_{+}}{2q q_{0}}\right)
+\mathcal{O}\left(\frac{1}{r^{2}}\right)\, ,
\\
& \nonumber\\
\mathcal{Z}_{0} 
& = 
1+\frac{1}{r}
\left[q_{0}+\frac{\alpha'}{2R}\left(-\frac{2}{n}+N_{\lambda} \mathfrak{n}_{\lambda}\right)\right]
+\mathcal{O}\left(\frac{1}{r^{2}}\right)\, ,
\end{align}

\noindent
where $\mathfrak{n}_\lambda=1/n$ is the instanton number, as given in
Eq.~(\ref{eq:instnumber}). Therefore, there are additional
$\mathcal{O}(\alpha')$ contributions to the $S5$ and $W$ charges at infinity,
while the $F1$ charge does not receive any contributions at
$\mathcal{O}(\alpha')$.

Let us first consider the corrections/contributions to the S5 charge through
the functions $\mathcal{Z}_{0}$. Some of these are immediately identified as
produced by gauge 5-branes, which play the r\^ole of non-localized S5
sources. They are responsible for the appearance of the instanton number in
the above expressions and are well understood.

The rest of the $\alpha'$ corrections in $\mathcal{Z}_{0}$ are associated to
the $\Omega_{(-)}$ connection. As we already remarked some lines above, their
net effect is the same as that of $N_{\lambda}=2$ gauge 5-branes but with a
negative sign. This means that we can eliminate all the $\alpha'$ corrections
in $\mathcal{Z}_{0}$ simply by taking $N_{\lambda}=2$. This choice is that of
the \emph{symmetric} 5-brane of Ref.~\cite{Callan:1991at} adapted to include a
KK monopole of charge $n$.

The case of S5 charge is paradigmatic, because we can explicitly identify and
compute the different types of sources; the pole in $\mathcal{Z}_{0}$ at $r=0$
gives the number of solitonic 5-branes, while the remaining regular functions
account for delocalized sources of non-Abelian nature coming from
$\alpha'$-corrections. Furthermore, these corrections can be completely
cancelled by an adequate choice of Yang-Mills fields eliminating any possible
ambiguity in the identification of the number of S5-branes that source the
solution.

Now we turn back our attention to the W charge and the interpretation of
$q_{+}$. At this stage and after the preceding discussion, the direct 
identification of $q_{+}$ in terms of $N_{W}$ units of momentum charge carried
by the string proposed in Eq.~(\ref{eq:charges}) is rather natural; the pole
in $\mathcal{Z}_{+}$ should correspond to a localized source of momentum,
while the remaining regular contribution to the function represents
delocalized sources of momentum arising from higher-order interactions. The
additional contributions to $N_{W}$ at infinity coming from the $\alpha'$
corrections have positive sign. We do not know how to cancel them by
introducing some delocalized gauge brane. The contribution of a
$\mathrm{SU}(2)$ dyonic gauge field, like that of Ref.~\cite{Ramirez:2016tqc},
would only add up to it. This is an $\alpha'$ correction that we have to live
with and which has important physical consequences.

Performing a T-duality transformation along the $u$ coordinate using the
$\alpha'$-corrected rules \cite{Bergshoeff:1995cg} amounts to the following
changes in the functions \cite{Cano:2018qev, Chimento:2018kop}

\begin{align}
\mathcal{Z}_{-} & \rightarrow \mathcal{Z}'_{-} = 1+\frac{1}{r}\left(q_{+} + \frac{\alpha'q_{+}}{2q q_{0}} \right) \, , 
\nonumber \\ 
& \\
\mathcal{Z}_{+} & \rightarrow \mathcal{Z}'_{+}= 1+ \frac{1}{r}\left(q_{-} - \frac{\alpha'q_{-}}{2q q_{0}} \right) + 
\frac{\alpha' q_{-}}{2qq_{0}} 
\frac{r^{2}+r(q_{0}+q_{+}+q)+q q_{0}+q q_{+}+q_{0}q_{+}
}{(r+q)(r+q_{0})(r+q_{+})} \, .
\nonumber 
\end{align}

\noindent
As we see, T-duality interchanges the total (asymptotic) string and momentum
charges.\footnote{On the other hand, the near-horizon geometry and entropy are preserved at $\mathcal{O}(\alpha')$. Moreover, temperature and entropy of the BTZ black hole in a simplified model remain invariant under the $\alpha'$-corrected T-duality transformation, \cite{Edelstein:2018ewc}.} However, since the total momentum 
charge receives contributions from
delocalized sources while the total string charge does not, the microscopic
T-duality rules interchanging $N_{W}$ and $N_{F1}$ have to be modified as

\begin{align}
q'_{-} &=\left(q_{+} + \frac{\alpha'q_{+}}{2q q_{0}} \right)  &  \longrightarrow
\hspace{1.5cm} & N'_{F1}= N_{W} \left( 1+ \frac{2}{n N_{S5}} \right) \, , \\
& & \nonumber \\
q'_{+} &=\left(q_{-} - \frac{\alpha'q_{-}}{2q q_{0}} \right) & \longrightarrow \hspace{1.5cm} & N'_{W}= N_{F1} \left( 1- \frac{2}{nN_{S5}} \right) \, .
\end{align}

We emphasize that the identification of the parameters of the solution in
terms of fundamental String Theory objects and their relation with the
asymptotic charges is affected by the global aspects of the solution when
$\alpha'$ corrections are present. Therefore, the use of merely near-horizon
geometries for this purpose, which, as we have discussed in the Introduction,
has been a common strategy in the literature, can introduce errors in the
determination of the sources. We shall come back to this issue in the
discussion section.

To close this section, we note that when one or both of $q,q_{0}$ vanish, a
qualitatively different family of solutions, known as ``small'' black holes,
is obtained. In order to avoid difficulties in the limits $q, q_{0}\rightarrow
0$ in $\mathcal{Z}_{0}$ given by Eq.~(\ref{eq:exp_Z02}) it is convenient to
work with the symmetric solution with $N_{\lambda}=2$ gauge 5-branes that
kills the $\alpha'$ corrections. On the other hand, this limit is singular in
the expression of $\mathcal{Z}_{+}$ given in Eq.~(\ref{eq:exp_Z+2}), but the
reason is that the subtraction made in Eq.~(\ref{eq:subtractedZs}) is not
pertinent in this case. Hence, we should use Eq.~(\ref{eq:exp_Z+}). For
example, in the $q_{0}\rightarrow 0$ limit, we get

\begin{align}
\label{eq:exp_Z+3}
\mathcal{Z}_{+} 
& = 
1+\frac{q_{+}}{r}-\frac{\alpha' q_{+} q_{-} }{2r^{2} (r+q)(r+q_{-})}
+\mathcal{O}(\alpha'{}^{2})\, ,
\\
& \nonumber\\
\label{eq:exp_Z-3}
\mathcal{Z}_{-}  
& = 
1+\frac{q_{-}}{r}+\mathcal{O}(\alpha'{}^{2})\, ,
\\
& \nonumber\\
\label{eq:exp_Z03}
\mathcal{Z}_{0}  
& = 
1\, ,\\
& \nonumber\\
\label{eq:exp_H3}
\mathcal{H}
& = 
1+\frac{q}{r}+\mathcal{O}(\alpha'{}^{2})\, . 
\end{align}

An analogous solution is obtained if instead we set $q=0$, $q_{0}\neq 0$, just
by interchanging $\mathcal{H}$ with $\mathcal{Z}_{0}$ and $q_{0}$ with $q$ in
$\mathcal{Z}_{+}$. These solutions are characterized by the anomalous degree
of divergence of $\mathcal{Z}_{+}$ at $r\rightarrow 0$, which now is $\sim
1/r^{2}$, instead of the usual $\sim 1/r$.  If both $q$ and $q_{0}$ vanish,
the divergence is $\sim 1/r^{3}$. As we discuss in the next section, this
behavior potentially could stretch the horizon of these small black holes to
yield a non-vanishing area. However, we will see in Section~\ref{sec:small}
that these solutions are non-perturbative and therefore the functions given in
Eqs.~(\ref{eq:exp_Z+3})-(\ref{eq:exp_H3}) cannot give a good description of
the small black holes near the horizon.

%%%%%%%%%%%%%%%%%%%%%%%%%%%%%%%%%%%%%%%%%%%%%%%%%%%%%%%%%%%%%%%%%%%%%%
%%%%%%%%%%%%%%%%%%%%%%%%%%%%%%%%%%%%%%%%%%%%%%%%%%%%%%%%%%%%%%%%%%%%%%
%%%%%%%%%%%%%%%%%%%%%%%%%%%%%%%%%%%%%%%%%%%%%%%%%%%%%%%%%%%%%%%%%%%%%%
%%%%%%%%%%%%%%%%%%%%%%%%%%%%%%%%%%%%%%%%%%%%%%%%%%%%%%%%%%%%%%%%%%%%%%
\section{\texorpdfstring{$\alpha'$}{α'} corrections to 
\texorpdfstring{$d=4$}{d=4} black holes}
\label{sec-4d}
%%%%%%%%%%%%%%%%%%%%%%%%%%%%%%%%%%%%%%%%%%%%%%%%%%%%%%%%%%%%%%%%%%%%%%
%%%%%%%%%%%%%%%%%%%%%%%%%%%%%%%%%%%%%%%%%%%%%%%%%%%%%%%%%%%%%%%%%%%%%%
%%%%%%%%%%%%%%%%%%%%%%%%%%%%%%%%%%%%%%%%%%%%%%%%%%%%%%%%%%%%%%%%%%%%%%
%%%%%%%%%%%%%%%%%%%%%%%%%%%%%%%%%%%%%%%%%%%%%%%%%%%%%%%%%%%%%%%%%%%%%%

From the solutions constructed in the previous section and the
compactification given by Eq.~(\ref{eq:4chargebh}), we obtain a family of
4-dimensional black holes with $\alpha'$ corrections. Let us determine the
main properties of these solutions.

%%%%%%%%%%%%%%%%%%%%%%%%%%%%%%%%%%%%%%%%%%%%%%%%%%%%%%%%%%%%%%%%%%%%%%
%%%%%%%%%%%%%%%%%%%%%%%%%%%%%%%%%%%%%%%%%%%%%%%%%%%%%%%%%%%%%%%%%%%%%%
%%%%%%%%%%%%%%%%%%%%%%%%%%%%%%%%%%%%%%%%%%%%%%%%%%%%%%%%%%%%%%%%%%%%%%
%%%%%%%%%%%%%%%%%%%%%%%%%%%%%%%%%%%%%%%%%%%%%%%%%%%%%%%%%%%%%%%%%%%%%%
\subsection{4-charge black holes}
%%%%%%%%%%%%%%%%%%%%%%%%%%%%%%%%%%%%%%%%%%%%%%%%%%%%%%%%%%%%%%%%%%%%%%
%%%%%%%%%%%%%%%%%%%%%%%%%%%%%%%%%%%%%%%%%%%%%%%%%%%%%%%%%%%%%%%%%%%%%%
%%%%%%%%%%%%%%%%%%%%%%%%%%%%%%%%%%%%%%%%%%%%%%%%%%%%%%%%%%%%%%%%%%%%%%
%%%%%%%%%%%%%%%%%%%%%%%%%%%%%%%%%%%%%%%%%%%%%%%%%%%%%%%%%%%%%%%%%%%%%%

Let us first consider the case in which all charges are non-vanishing, $q
q_{0} q_{+} q_{-}\neq 0$.  We write here the metric of these black holes for
convenience,

\begin{equation}
ds^{2}  
= 
e^{2U}dt^{2}-e^{-2U}d\vec{x}^{\, 2}\, ,
\quad 
\text{where}\,\,\,\, 
e^{-2U} = \sqrt{\mathcal{Z}_{0}\, \mathcal{Z}_{+}\, \mathcal{Z}_{-} \mathcal{H}}\, .
\end{equation}

\noindent
First, defining the mass $M$ from the asymptotic behavior

\begin{equation}
e^{-2U}=1+\frac{2 G_{N}^{(4)}M}{r}+\mathcal{O}\left(\frac{1}{r^{2}}\right)\, ,
\end{equation}

\noindent
and using Eqs.~(\ref{eq:exp_Z0})-(\ref{eq:exp_H}), we obtain

\begin{equation}
\label{M1}
M
=
\frac{1}{4 G_{N}^{(4)}}
\left[q_{0}+\frac{\alpha'(N_{\lambda}-2)}{4q}+q_{-}+q_{+}
\left(1+\frac{\alpha'}{2q q_{0}}\right)+q\right]\, .
\end{equation}

This expression takes a more meaningful form in terms of the stringy objects
that form the black hole.  The charges are characterized by the integer
numbers $N_{S5}$, $N_{F1}$, $N_{W}$ and $n$ according to Eqs.~(\ref{eq:charges})
and (\ref{quant}). On the other hand, the 4-dimensional Newton's constant is
obtained from the 10-dimensional one as

\begin{equation}
\label{eq:Newtonconstant4d}
G_{N}^{(4)}
=
\frac{G_{N}^{(10)}}{(2\pi R)(2\pi R_{u})(2\pi \ell_{s})^{4}}
=
\frac{g_{s}^{2}\ell_{s}^{4}}{8 R R_{u}}\, ,
\end{equation}

\noindent
where we used the value of the 10-dimensional Newton's constant
$G_{N}^{(10)}=8\pi^{6}\ell_{s}^{8} g_{s}^{2}$.  Plugging this into
Eq.~(\ref{M1}), we get

\begin{equation}\label{M2}
M
=
\frac{R_{u}}{g_{s}^{2}\ell_{s}^{2}}\left(N_{S5}+\frac{N_{\lambda}-2}{n}\right)
+\frac{R_{u}}{\ell_{s}^{2}}N_{F1}
+\frac{N_{W}}{R_{u}}\left(1+\frac{2}{n N_{S5}}\right)
+n\frac{R^{2} R_{u}}{g_{s}^{2}\ell_{s}^{4}} \, .
\end{equation}

We see that every $\mathrm{SU}(2)$ instanton contributes to the mass as one
S5-brane times the instanton number.\footnote{In our previous works
  (Refs.~\cite{Cano:2018qev,Chimento:2018kop,Cano:2017qrq,Cano:2017sqy,Cano:2017nwo})
  we have used, following Ref.~\cite{Strominger:1990et}, a (wrong)
  normalization that differs from the one used in this paper by a factor of 8
  for the $\alpha'$ corrections in the action, so in that references every
  instanton contributes as 8 S5 branes. Here we have eliminated this factor,
  in agreement with Ref.~\cite{Duff:1994an}. We thank Prof.~M.J.~Duff for
  sharing this information with us.} If we do not include $\mathrm{SU}(2)$
fields, the $\alpha'$ corrections induce a negative mass term that looks as
that of certain kind of anti S5-brane.  Hence, in order to avoid negative mass
contributions the most natural choice corresponds to $N_{\lambda}=2$, and in
particular to the choice of two instantons that cancel all the $\alpha'$
corrections in the function sourced by the 5-branes $\mathcal{Z}_{0}$. Then,
the S5-branes are symmetric.

An intriguing observation in the symmetric case is the following. As we see,
the mass is the sum of four terms, that actually correspond to the asymptotic
charges as should happen in a BPS state.  Let us write it as
$M=Q_{1}+Q_{2}+Q_{2}+Q_{4}$. Then, if we consider instead the quantity

\begin{equation}
16 \pi G_{N}^{(4)} \sqrt{Q_{1}Q_{2}Q_{3}Q_{4}}=2\pi\sqrt{N_{F1}N_{W}(n N_{S5}+2)}\, ,
\end{equation}

\noindent
we are going to see that it coincides with the exact result for the entropy
computed counting string microstates. At the very least, this relation
suggests that in the symmetric case we do not expect to have further
corrections to the mass, and hence the resulting Eq.~(\ref{M2}) with
$N_{\lambda}=2$ would be exact. Going further, this also supports the
conjecture, based on the form of the $T$-tensors that account for all the
$\alpha'$ corrections associated to the supersymmetrization of the
Chern-Simons terms, that the symmetric solution could be actually exact at all
orders in $\alpha'$.\footnote{The reasons that support this conjecture are the
  exactly the same that support it in the 5-dimensional case treated in
  Ref.~\cite{Cano:2018qev}. Unfortunately, the $\alpha'$ corrections which are
  unrelated to the supersymmetrization of the Chern-Simons terms are not well
  known and there is no much that can be said about them that supports or
  contradicts the conjecture.}

On the other hand, the near-horizon geometry $r\rightarrow 0$ is that of
$\mathrm{AdS_{2}}\times \mathrm{S}^{2}$ and it does not receive any explicit
$\alpha'$ correction:

\begin{equation}
ds_{\rm{nh}}^{2}
=
(q_{0} q_{+}q_{-}q)^{-1/2} r^{2} dt^{2}+(q_{0}
q_{+}q_{-}q)^{1/2}\left(\frac{dr^{2}}{r^{2}}
+d\Omega_{(2)}^{2}\right)\, ,
\end{equation}

\noindent
and the area of the horizon area is 

\begin{equation}
\label{A1}
A=4\pi \sqrt{q_{0} q_{+}q_{-} q}\, .
\end{equation}

\noindent
However, due to the terms of hiher order in the curvature present in the
Heterotic Superstring effective action, the entropy is not simply given by
$A/(4 G_{N}^{(4)})$. We will compute its value in the next section by applying
Wald's formula \cite{Wald:1993nt,Iyer:1994ys}.

%%%%%%%%%%%%%%%%%%%%%%%%%%%%%%%%%%%%%%%%%%%%%%%%%%%%%%%%%%%%%%%%%%%%%%
%%%%%%%%%%%%%%%%%%%%%%%%%%%%%%%%%%%%%%%%%%%%%%%%%%%%%%%%%%%%%%%%%%%%%%
%%%%%%%%%%%%%%%%%%%%%%%%%%%%%%%%%%%%%%%%%%%%%%%%%%%%%%%%%%%%%%%%%%%%%%
%%%%%%%%%%%%%%%%%%%%%%%%%%%%%%%%%%%%%%%%%%%%%%%%%%%%%%%%%%%%%%%%%%%%%%
\subsubsection*{Bonus example: corrections to Reissner-Nordstrom black hole}
%%%%%%%%%%%%%%%%%%%%%%%%%%%%%%%%%%%%%%%%%%%%%%%%%%%%%%%%%%%%%%%%%%%%%%
%%%%%%%%%%%%%%%%%%%%%%%%%%%%%%%%%%%%%%%%%%%%%%%%%%%%%%%%%%%%%%%%%%%%%%
%%%%%%%%%%%%%%%%%%%%%%%%%%%%%%%%%%%%%%%%%%%%%%%%%%%%%%%%%%%%%%%%%%%%%%
%%%%%%%%%%%%%%%%%%%%%%%%%%%%%%%%%%%%%%%%%%%%%%%%%%%%%%%%%%%%%%%%%%%%%%

The Reissner-Nordstr\"om black hole corresponds to the zeroth-order in
$\alpha'$ solution with $q_{+}=q_{-}=q_{0}=q$. This choice of charges gives
constant scalars $e^{\phi}=e^{\phi_{\infty}},k=k_{\infty}$ and
$\ell=\ell_{\infty}$ at this order. However, taking into account the
constituents of the black hole, we can only take those charges equal at given
points in moduli space

\begin{equation}
g_{s}=e^{\phi_{\infty}}=\sqrt{\frac{N_{S5}}{N_{F1}}}\, ,
\hspace{1cm}
R_{z}/\ell_{s} = k_{\infty} = \sqrt{\frac{N_{W}}{N_{F1}}}\, ,  
\hspace{1cm}
R/\ell_{s} = 2\ell_{\infty} = \sqrt{\frac{N_{S5}}{n}}\, ,  
\end{equation}

\noindent
which fixes the asymptotic values of the scalars to their attractor values.

Applying the general result to this particular case is straightforward. Taking
the symmetric case, we find that the corrected metric function $e^{-2U}$ and
the scalars take the form

\begin{eqnarray}
e^{-2U}
& = & 
\left(1+\frac{q}{r}\right)^{2}
+\frac{\alpha'}{4q}\frac{[r^{2}+3rq+3q^{2}]}{r(r+q)^{2}} +\mathcal{O}(\alpha'{}^{2})\, ,
\\
& & \nonumber \\
e^{2\phi}
& = &
e^{2\phi_{\infty}} \, ,
\\
& & \nonumber \\
k
& = & 
k_{\infty} +\frac{\alpha' k_{\infty}}{4q}
\frac{r[r^{2}+3rq+3q^{2}]}{(r+q)^{4}}
 +\mathcal{O}(\alpha'{}^{2})\, ,
\\
& & \nonumber \\
\ell
& = & 
\ell_{\infty} +\frac{\alpha' \ell_{\infty}}{12q}
\frac{r[r^{2}+3rq+3q^{2}]}{(r+q)^{4}}
 +\mathcal{O}(\alpha'{}^{2})\, ,
\end{eqnarray}

\noindent
with

\begin{equation}
q = \frac{\ell_{s}}{2} \sqrt{nN_{S5}}\, .
\end{equation}

We also have to take into account that the 4-dimensional Newton constant given
in Eq.~(\ref{eq:Newtonconstant4d}) now has the value

\begin{equation}
G_{N}^{(4)}
=
\frac{\ell^{2}_{s}}{8}\sqrt{\frac{nN_{S5}}{N_{W}N_{F1}}}\, .  
\end{equation}

\noindent
Then, if we do not want $G^{(4)}_{N}$ to change with $q$, we must set
$N_{W}N_{F1}=\aleph^{2}N_{S5}n$ for some positive dimensionless constant
$\aleph$ so that

\begin{equation}
G_{N}^{(4)}
=
\frac{\ell^{2}_{s}}{8 \aleph}\, .
\end{equation}

Staying at weak coupling (so the loop corrections can be safely ignored) and away
of the self-dual radii at which new massless degrees of freedom arise, demands
the following hierarchy 

\begin{equation}
N_{W} >N_{F1} > N_{S5} >n\, ,
\,\,\,\,\,\,
\Rightarrow
\,\,\,\,\,\,
\aleph >>1\, .  
\end{equation}

At this point in moduli space, the mass of the black hole will be given by

\begin{equation}
M
= 
\frac{4}{\ell_{s}}\sqrt{N_{W}N_{F1}}\left(1+\frac{1}{2nN_{S5}}\right) \, ,
%=
%\frac{4\aleph}{\ell_{s}}\left(\sqrt{nN_{S5}} +\frac{4}{\sqrt{nN_{S5}}}\right)\, ,
\,\,\,\,\,
\text{or}
\,\,\,\,\,
2G^{(4)}_{N}M 
= 
\ell_{s} \left(\sqrt{nN_{S5}} +\frac{1}{2\sqrt{nN_{S5}}}\right)\, ,   
\end{equation}

\noindent
while the area of the horizon takes the value

\begin{equation}
A=4\pi q^{2} =\pi\ell_{s}^{2} nN_{S5}\, ,  
\end{equation}

\noindent
and the leading contribution to the entropy will be

\begin{equation}
\frac{A}{4G^{(4)}_{N}} = 2 \pi \sqrt{nN_{S5}N_{W}N_{F1}} = 2 \pi \aleph
nN_{S5}\, . 
\end{equation}

%%%%%%%%%%%%%%%%%%%%%%%%%%%%%%%%%%%%%%%%%%%%%%%%%%%%%%%%%%%%%%%%%%%%%%
%%%%%%%%%%%%%%%%%%%%%%%%%%%%%%%%%%%%%%%%%%%%%%%%%%%%%%%%%%%%%%%%%%%%%%
%%%%%%%%%%%%%%%%%%%%%%%%%%%%%%%%%%%%%%%%%%%%%%%%%%%%%%%%%%%%%%%%%%%%%%
%%%%%%%%%%%%%%%%%%%%%%%%%%%%%%%%%%%%%%%%%%%%%%%%%%%%%%%%%%%%%%%%%%%%%%
\subsection{Small black holes}
\label{sec:small}
%%%%%%%%%%%%%%%%%%%%%%%%%%%%%%%%%%%%%%%%%%%%%%%%%%%%%%%%%%%%%%%%%%%%%%
%%%%%%%%%%%%%%%%%%%%%%%%%%%%%%%%%%%%%%%%%%%%%%%%%%%%%%%%%%%%%%%%%%%%%%
%%%%%%%%%%%%%%%%%%%%%%%%%%%%%%%%%%%%%%%%%%%%%%%%%%%%%%%%%%%%%%%%%%%%%%
%%%%%%%%%%%%%%%%%%%%%%%%%%%%%%%%%%%%%%%%%%%%%%%%%%%%%%%%%%%%%%%%%%%%%%

It has been known for some time that, in the String Theory context and at
lowest order in $\alpha'$, four charges are needed in order to obtain a
4-dimensional extremal black hole with a regular horizon.  When one or more of
the charges vanish, the horizon (still located at $r=0$) has zero area and
becomes singular. Usually, the small black holes considered in the literature
only contain $q_{-}$ and $q_{+}$ charges (corresponding to strings and
waves). There is a curvature singularity at $r=0$ and the scalars behave there as

\begin{equation}
e^{2\phi}\rightarrow 0\, , 
\quad 
k\sim \frac{1}{r^{1/4}}\, , 
\quad 
\ell\sim \frac{1}{r^{1/3}}\, .
\end{equation}

\noindent
Although at zeroth-order in $\alpha'$ the area and entropy vanish, the exact
entropy of such black holes computed by counting string microstates is finite
(see e.g. Ref.~\cite{Sen:2007qy}). Hence, from the Supergravity perspective,
it was expected that the singularity at the would-be horizon of these
solutions could be fixed somehow. Since the string coupling vanishes at $r=0$,
quantum corrections cannot be of help here, but $\alpha'$ corrections can be
relevant because the curvature is very large (divergent) there.

Let us consider the $\alpha'$-corrected solution in the case $q_{0}= 0$,
allowing $q$ to be arbitrary, so that the solution still contains 3
independent charges. This solution is determined by the functions in
Eqs.~(\ref{eq:exp_Z+3})-(\ref{eq:exp_H3}). Let us recall that we are choosing
a symmetric KK monopole in this case, so that $\mathcal{Z}_{0}=1$.  This
solution is qualitatively different from the 4-charge black hole. For example,
the limit $q_{0}\rightarrow 0$ in the mass formula Eq.~(\ref{M1}) would be
divergent, but we find instead

\begin{equation}
\label{Ms}
M=\frac{1}{4 G_{N}^{(4)}}\left[q_{-}+q_{+}+q\right]\, .
\end{equation}

On the other hand, in the limit $r\rightarrow 0$ the functions $\mathcal{H}$
and $\mathcal{Z}_{-}$ diverge with the usual $1/r$ behavior, but the $\alpha'$
corrections produce another divergence in $\mathcal{Z}_{+}$ that dominates in
this limit:

\begin{equation}
\label{eq:Z+small}
\mathcal{Z}_{+}\sim -\frac{\alpha' q_{+}}{2 q r^{2}}\, .
\end{equation}

\noindent
Thus, assuming that $q_{+}q_{-}<0$, the metric becomes again
$\mathrm{AdS}_{2}\times \mathrm{S}^{2}$ and the area of the horizon is

\begin{equation}\label{A2}
A=4\pi \sqrt{-\alpha' q_{+} q_{-}/2}\, .
\end{equation}

Interestingly, only two charges contribute to this area even if the solution
contains three of them.  It is immediate to check that the previous formula also
holds if we set $q=0$ (because then $\mathcal{Z}_{+}\sim -\frac{\alpha'
  q_{+}}{2r^{3}}$) or if we set instead $q_{0}\neq0$, $q=0$. Summarizing all
the cases, if $q_{0}q=0$ it seems that the higher curvature corrections pump the area of the horizon up from zero to the value in Eq.~(\ref{A2}).

However, it is possible to see that this analysis is not rigorous enough. Our
solution generating technique, summarized in Section~\ref{sec-solutions},
assumes that the solution admits a perturbative expansion. In particular, to
obtain the corrections to the function $\mathcal{Z}_{+}$ as given in
Eq.~(\ref{eq:Z+}) we implicitly assume that the dominant contribution to
$\partial_{n} \mathcal{Z}_{+} $ is given by $\partial_{n}
\mathcal{Z}_{+}^{(0)}$ everywhere, allowing a perturbative construction. This
is true in the 4-charge family of solutions, but when $q_{0}q=0$ we
immediately see from Eq.~(\ref{eq:Z+small}) that $\partial_{n}
\mathcal{Z}_{+}^{(0)}$ is subdominant in the near-horizon region. This is a
contradiction which signals that our method to build solutions is not valid in
this case and that small black holes are non perturbative, so the solution
Eqs.~(\ref{eq:exp_Z+3})-(\ref{eq:exp_H3}) should not be trusted at this stage.
They will be studied in more detail in a forthcoming paper Ref.~\cite{kn:CRR}.\\

Let us close this section by mentioning that $\alpha'$ corrections also introduce very curious 
properties in 4-charge solutions when two of the charges are negative. At zeroth order in $\alpha'$
such solutions represent naked singularities, but it was shown in Ref.~\cite{Cano:2018aod} that the
first-order corrections turn them into globally regular black holes whose horizon does not contain any singularity.
These solutions suffer from the same issue as the one mentioned for small black holes,
hence a thorough analysis of the higher-order corrections is necessary to determine if these appealing 
regular black holes can be trusted.

%%%%%%%%%%%%%%%%%%%%%%%%%%%%%%%%%%%%%%%%%%%%%%%%%%%%%%%%%%%%%%%%%%%%%%
%%%%%%%%%%%%%%%%%%%%%%%%%%%%%%%%%%%%%%%%%%%%%%%%%%%%%%%%%%%%%%%%%%%%%%
%%%%%%%%%%%%%%%%%%%%%%%%%%%%%%%%%%%%%%%%%%%%%%%%%%%%%%%%%%%%%%%%%%%%%%
%%%%%%%%%%%%%%%%%%%%%%%%%%%%%%%%%%%%%%%%%%%%%%%%%%%%%%%%%%%%%%%%%%%%%%
\section{Black hole entropy}
\label{sec-entropy}
%%%%%%%%%%%%%%%%%%%%%%%%%%%%%%%%%%%%%%%%%%%%%%%%%%%%%%%%%%%%%%%%%%%%%%
%%%%%%%%%%%%%%%%%%%%%%%%%%%%%%%%%%%%%%%%%%%%%%%%%%%%%%%%%%%%%%%%%%%%%%
%%%%%%%%%%%%%%%%%%%%%%%%%%%%%%%%%%%%%%%%%%%%%%%%%%%%%%%%%%%%%%%%%%%%%%
%%%%%%%%%%%%%%%%%%%%%%%%%%%%%%%%%%%%%%%%%%%%%%%%%%%%%%%%%%%%%%%%%%%%%%

In this section we will compute the Wald entropy of the black holes we have
obtained, up to terms of $\mathcal{O}(\alpha')$.  Following
Refs.~\cite{Wald:1993nt,Iyer:1994ys}, the Wald entropy formula for a
$D+1$-dimensional theory is

\begin{equation}
 \mathbb{S}
=
-2\pi \int_\Sigma d^{D-1}x\sqrt{|h|}\mathcal{E}_{R}^{abcd}\epsilon_{ab}\epsilon_{cd}\,,
\end{equation}

\noindent
where $\Sigma$ is a cross section of the horizon, $h$ is the determinant of
the metric induced on $\Sigma$, $\epsilon_{ab}$ is the binormal to $\Sigma$
with normalization $\epsilon_{ab}\epsilon^{ab}=-2$, and
$\mathcal{E}_{R}^{abcd}$ is the equation of motion one would obtain for the
Riemann tensor $R_{abcd}$ treating it as an independent field of the theory,

\begin{equation}
\mathcal{E}_{R}^{abcd} 
= 
\frac{1}{\sqrt{|g|}}\frac{\delta S_{(D+1)}}{\delta R_{abcd}}\, ,
\end{equation}

\noindent
where $S_{(D+1)}$ is the action of the theory.

Then it seems that, in order to compute the entropy of the lower dimensional
solutions one needs to know the $\alpha'$ corrections to the Lagrangians of
the dimensionally reduced theories. The determination of such correction terms
starting from the 10-dimensional theory would require a long and intricate
calculation, but luckily it turns out to be unnecessary since, as we will now
show following the discussion in Ref.~\cite{Cano:2018qev}, the entropy of the
$4$- and $5$-dimensional solutions can be re-expressed entirely in terms of
integrals of $10$-dimensional quantities.

It is convenient to work in frames adapted to the dimensional reduction. To
this end, we start by rewriting the $10$- and $5$-dimensional line elements
Eqs.~(\ref{eq:metric}) and (\ref{eq:3chargebh}) in the following form:

\begin{align}
ds_{(10)}^{2}  
& = 
e^{\phi-\phi_\infty}\left[ (k/k_\infty)^{-2/3}ds_{(5)}^{2}
-(k/k_\infty)^{2} \left( du-\frac{dt}{\mathcal{Z}_{+}} \right)^{2}\right]
-dy^{i} dy^{i}\, ,
\\
\nonumber\\
ds_{(5)}^{2} 
& = 
R/(2\ell)\, ds_{(4)}^{2}-\ell^{2} \left( d\Psi+\chi \right)^{2}\, ,
\end{align}

\noindent
where the $4$-dimensional line element $ds_{(4)}^{2}$, the dilaton $\phi$ and
the Kaluza-Klein scalars $k$ and $\ell$ are given by (\ref{eq:4chargebh}),
which we report here for convenience:

\begin{equation}
\begin{array}{rcl}
ds_{(4)}^{2} 
& = & 
e^{2U}dt^{2}-e^{-2U}d\vec{x}^{\, 2}\, ,
\\
& & \\
e^{2\phi}
& = &
e^{2\phi_{\infty}}{\displaystyle\frac{\mathcal{Z}_{0}}{\mathcal{Z}_{-}}}\, ,
\hspace{1cm}
k
= 
k_{\infty} 
{\displaystyle\frac{\mathcal{Z}_{+}^{1/2}}{\mathcal{Z}_{0}^{1/4}\mathcal{Z}_{-}^{1/4}}}\, ,
\hspace{1cm}
\ell
= 
\ell_{\infty} 
{\displaystyle\frac{\mathcal{Z}_{0}^{1/6}\, \mathcal{Z}_{+}^{1/6}\,
    \mathcal{Z}_{-}^{1/6}}{\mathcal{H}^{1/2}}}\, , 
\label{eq:kkscalars}
\end{array}
\end{equation}

\noindent
with $e^{\phi_{\infty}}=g_{s}$ and 

\begin{equation}
e^{-2U} = \sqrt{\mathcal{Z}_{0}\, \mathcal{Z}_{+}\, \mathcal{Z}_{-} \mathcal{H}}\, .
\end{equation}

\noindent
We are also going to need the metric function of the 5-dimensional black holes
$f$, which is given in Eq.~(\ref{eq:f}) and which we also quote here for
convenience:

\begin{equation}
f^{-3}
=
\mathcal{Z}_{0}\, \mathcal{Z}_{+}\, \mathcal{Z}_{-}\, .
\end{equation}

We define the following Vielbein bases in $4$, $5$ and $10$ dimensions,
respectively:

\begin{gather}
 v^{0} = e^{U}  dt\,,\qquad v^{1}=e^{-U}  dr\,,\qquad v^{2} = e^{-U} r d\theta\,,
 \qquad v^{3} = e^{-U}r \sin\theta d\phi\, .
\\
\nonumber\\
V^{0,\ldots,3} = (R/(2\ell))^{1/2}\, v^{0,\ldots,3}\, ,
\qquad 
V^{4} = \ell \left( d\Psi+\cos{\theta} d\phi \right)\, ,
\\
\nonumber\\
W^{0,\ldots,4} = e^{\frac{\phi-\phi_\infty}{2}}(\frac{k}{k_\infty})^{-\frac13}
V^{0,\ldots,4}\, ,\;\;
W^{5} = e^{\frac{\phi-\phi_\infty}{2}} \frac{k}{k_\infty}
 \left( du-\frac{dt}{\mathcal{Z}_{+}} \right)\, , \;\;
W^{6,\ldots,9}=dy^{6,\ldots,9}\,,
\end{gather}

\noindent
where $(r,\theta,\phi)$ are spherical coordinates on $\mathbb{E}^{3}$.

As we can see, the Vielbein 1-form bases we have introduced in 4,5 and 10 dimensions
are related by multiplication by a common factor, and consequently the
components of the Riemann tensors in these frames are related by

\begin{equation}
 R_{(10)abcd} = e^{-(\phi-\phi_\infty)}(\frac{k}{k_\infty})^{2/3} R_{(5)abcd}+\ldots\,,
\end{equation}

\noindent
for $a,b,c,d=0,\ldots,4$ and

\begin{equation}
R_{(5)abcd} 
= 
\frac{2\ell}{R} \, R_{(4)abcd}+\ldots
\quad\Rightarrow\quad 
R_{(10)abcd} =
\frac{2\ell}{R}\,e^{-(\phi-\phi_\infty)}(\frac{k}{k_\infty})^{2/3} 
R_{(4)abcd}+\ldots\, ,
\end{equation}

\noindent
for $a,b,c,d=0,\ldots,3$, so that

\begin{equation}
\frac{\delta S_{(10)}}{\delta R_{(5)abcd}}
=
e^{-(\phi-\phi_\infty)}(k/k_\infty)^{2/3}
\int\frac{\delta S_{(10)}}{\delta R_{(10)abcd}}\, ,
\end{equation}

\noindent
and

\begin{equation}
 \frac{\delta S_{(10)}}{\delta R_{(4)abcd}}
=
\frac{2\ell}{R}\,e^{-(\phi-\phi_\infty)}(k/k_\infty)^{2/3}
\int\frac{\delta S_{(10)}}{\delta R_{(10)abcd}}\, ,
\end{equation}

\noindent
where the integrations are on the appropriate compact coordinates.
The action is, of course, the same in any dimension, and, therefore,

\begin{equation}
\frac{\delta S_{(5)}}{\delta R_{(5)abcd}}
=
\frac{\delta S_{(10)}}{\delta R_{(5)abcd}}\, ,
\qquad 
\frac{\delta S_{(4)}}{\delta R_{(4)abcd}}
= 
\frac{\delta S_{(10)}}{\delta R_{(4)abcd}}\, .
\end{equation}

In the solutions we are interested in, the horizon $\Sigma$ is located at
$r=0$, where the timelike Killing vector becomes null. Then we have 

\begin{equation}
 |g_{(5)}| = f \mathcal{H} |h_{(5)}|\, ,
\qquad 
|g_{(4)}| = |h_{(4)}|\, ,
\end{equation}

\noindent
on $\Sigma$, and the components of the binormal $\epsilon_{ab}$ when expressed
in the frames defined above are the same in any dimension, $\epsilon_{01}=1$.

All this allows us to write the Wald entropy for the $4$ and $5$-dimensional
solutions in $10$-dimensional language, the explicit expressions being

\begin{multline}
\mathbb{S}_{(5)} 
= 
-2\pi \int_{\Sigma_{3}} d^{3}x
\sqrt{\left|\frac{h_{(5)}}{g_{(5)}}\right|}
\frac{\delta S_{(5)}}{\delta  R_{(5)abcd}}\epsilon_{ab}\epsilon_{cd}
\\
\\
= 
-2\pi \int_{\Sigma_{3}\times S^{1}\times \mathrm{T}^{4}} d^{8} x
(f \mathcal{H})^{-1/2}e^{-(\phi-\phi_\infty)}(k/k_\infty)^{2/3}
\frac{\delta S_{(10)}}{\delta R_{(10)abcd}}\epsilon_{ab}\epsilon_{cd}\, ,
\end{multline}

\noindent
and

\begin{multline}
\mathbb{S}_{(4)} 
= 
-2\pi \int_{\Sigma_{2}} d^{2} x\sqrt{\left|\frac{h_{(4)}}{g_{(4)}}\right|}
\frac{\delta S_{(4)}}{\delta R_{(4)abcd}}\epsilon_{ab}\epsilon_{cd}
\\
\\
= 
-2\pi \int_{\Sigma_{2}\times S^{1}\times S^{1}\times \mathrm{T}^{4}} d^{8} x
\frac{2\ell}{R}\, e^{-(\phi-\phi_\infty)}(k/k_\infty)^{2/3}
\frac{\delta S_{(10)}}{\delta R_{(10)abcd}}\epsilon_{ab}\epsilon_{cd}\, .
\end{multline}

Observe that, taking into account the expression for $\ell$ in
Eq.~(\ref{eq:kkscalars}), if $\Sigma_{3} =\Sigma_{2}\times \mathrm{S}^{1}$,
then $\mathbb{S}_{(5)}=\mathbb{S}_{(4)}$.

The action $S_{(10)}$ is given in Eq.~(\ref{heterotic}).  At leading order it
depends on the Riemann tensor only through the Einstein-Hilbert term, while at
first order in $\alpha'$ there are additional contributions from terms
depending on the curvature of the torsionful spin connection $\Omega_{(-)}$,
denoted by $R_{(-)}$.  Since we are only interested in the first order
corrections and $R_{(-)}$ already appears at first order in the Lagrangian, it
is enough to consider the leading order dependence of $R_{(-)}$ on the Riemann
tensor,

\begin{equation}
 R_{(-)abcd}=R_{(10)abcd}+\ldots\, .
\end{equation}

\noindent
Since at this order no derivatives of the Riemann tensor appear in the
Lagrangian, one has on $\Sigma$

\begin{align}
\frac{1}{\sqrt{|g_{(10)}|}}\frac{\delta S_{(10)}}{\delta R_{(10)abcd}} 
\epsilon_{ab}\epsilon_{cd}
& = 
\frac{e^{-2(\phi-\phi_\infty)}}{16\pi G_{N}^{(10)}}
\frac{\partial}{\partial R_{(10)abcd}}\left(R+\frac{1}{2\cdot 3!}
  H^{2}-\frac{\alpha'}{8}  R_{(-)}{}^{a}{}_{b}
  R_{(-)}{}^{b}{}_{a}\right)\epsilon_{ab}\epsilon_{cd}
\nonumber \\
\nonumber \\
& =
\frac{e^{-2(\phi-\phi_\infty)}}{16\pi G_{N}^{(10)}} \left[\eta^{ac}\eta^{bd}+\frac{H^{efg}}{3!}  
\frac{\delta H_{efg}}{\delta R_{(10)abcd}}\right]\!\epsilon_{ab}\epsilon_{cd}
\nonumber \\
\nonumber \\
& = 
\frac{e^{-2(\phi-\phi_\infty)}}{16\pi G_{N}^{(10)}} \left[\eta^{ac}\eta^{bd}-
  \frac{\alpha'}{8}H^{abg} \Omega_{(-)g}{}^{cd}\right]\!\epsilon_{ab}\epsilon_{cd}
\nonumber\\
\nonumber\\ 
\label{eq:entropyH}
& =
-\frac{e^{-2(\phi-\phi_\infty)}}{8\pi G_{N}^{(10)}}
\left[1+ \frac{\alpha'}{4} H^{01g} \Omega_{(-)g}{}^{01}\right]\, .
\end{align}

The term quadratic in $R_{(-)}$ does not contribute because $R_{(-)}$ vanishes
on $\Sigma$.\footnote{See \cite{Prester:2008iu,Cano:2018qev}.} The relevant
components of the Kalb-Ramond field strength $H$ and of the torsionful spin
connection $\Omega_{(-)}$ can be obtained straightforwardly,

\begin{align}
H_{01g}
& =
-\delta_{g}^{5}(\mathcal{Z}_{0} \mathcal{H})^{-1/2} \partial_{r}\log{\mathcal{Z}_{-}}\, ,
\\
\nonumber\\
\Omega_{(-)501} 
& =
\tfrac{1}{2} (\mathcal{Z}_{0} \mathcal{H})^{-1/2} \partial_{r}
\log{(\mathcal{Z}_{-}\mathcal{Z}_{+})}\, ,
\\
\nonumber\\
H^{01g} \Omega_{(-)g}{}^{01} 
& = 
H^{015} \Omega_{(-)5}{}^{01}
= 
\tfrac{1}{2} \frac{\partial_{r}\log{(\mathcal{Z}_{-}\mathcal{Z}_{+})} 
\partial_{r}\log{\mathcal{Z}_{-}}}{\mathcal{Z}_{0} \mathcal{H}}\,. 
\end{align}

The last missing piece is the determinant of the 10-dimensional metric, which reads

\begin{equation}
 \sqrt{|g_{(10)}|}
=
e^{3(\phi-\phi_\infty)}f^{-1} \frac{R}{2}\, r^{2} \mathcal{H} \sin{\theta} (k/k_\infty)^{-2/3}\,.
\end{equation}

Putting everything together we get to

\begin{equation}
\mathbb{S}_{(4)}
= 
\frac{R}{8 G_{N}^{(10)}}\int d\theta d\phi d\Psi dz d^{4}y r^{2} \sin{\theta} 
\sqrt{\mathcal{Z}_{0} \mathcal{Z}_{+} \mathcal{Z}_{-} \mathcal{H}} 
 \left( 1+\frac{\alpha'}{8}
   \frac{\partial_{r}\log{(\mathcal{Z}_{-}\mathcal{Z}_{+})} 
\partial_{r}\log{\mathcal{Z}_{-}}}{\mathcal{Z}_{0} \mathcal{H}}\right)\,.
\end{equation}

Once we substitute the explicit form of the functions $\mathcal{Z}_{0,\pm}$
given in Eqs.~(\ref{eq:harmonic_Z}) and $\mathcal{H}$ given in
Eq.~(\ref{eq:HforKKmonopole}) and integrate on the compact coordinates and on
$\Sigma$ we arrive at the result

\begin{equation}
\mathbb{S}_{(4)}
= 
\frac{\pi}{G_{N}^{(4)}} 
\sqrt{q_{0}\, q_{+} q_{-} q}\left( 1+\frac{\alpha'}{4q_{0} q} \right)\, ,
\end{equation}

\noindent
with the $4$-dimensional Newton constant given by

\begin{equation}
G_{N}^{(4)}=\frac{G_{N}^{(10)}}{(2\pi R)(2\pi R_u)(2\pi \ell_{s})^{4}}\, .
\end{equation}

\noindent
Finally, with the identifications Eqs.~(\ref{eq:charges}) and (\ref{quant})

\begin{equation}
q_{0}=\frac{\alpha'}{2 R} N_{S5}\, ,
\qquad 
q_{+} = \frac{ \alpha'{}^{2} g_{s}^{2}}{2 R R_u^{2}} N_{W}\, ,
\qquad 
q_{-} = \frac{\alpha' g_{s}^{2}}{2 R} N_{F1}\, ,
\qquad q = \frac{n R}{2}\, ,
\end{equation}

\noindent
and 

\begin{equation}
 G_{N}^{(10)}=8\pi^{6} \alpha'{}^{4} g_{s}^{2}\,,
\end{equation}

\noindent
the entropy can be finally rewritten as

\begin{equation}
\label{eq:entropy1}
\mathbb{S}_{(4)}
=
2\pi \sqrt{ N_{F1}N_{W}n N_{S5} } \left(1+\frac{1}{n N_{S5}}\right)\, .
\end{equation}

%%%%%%%%%%%%%%%%%%%%%%%%%%%%%%%%%%%%%%%%%%%%%%%%%%%%%%%%%%%%%%%%%%%%%%
%%%%%%%%%%%%%%%%%%%%%%%%%%%%%%%%%%%%%%%%%%%%%%%%%%%%%%%%%%%%%%%%%%%%%%
%%%%%%%%%%%%%%%%%%%%%%%%%%%%%%%%%%%%%%%%%%%%%%%%%%%%%%%%%%%%%%%%%%%%%%
%%%%%%%%%%%%%%%%%%%%%%%%%%%%%%%%%%%%%%%%%%%%%%%%%%%%%%%%%%%%%%%%%%%%%%
\section{Discussion}
\label{sec-discussion}
%%%%%%%%%%%%%%%%%%%%%%%%%%%%%%%%%%%%%%%%%%%%%%%%%%%%%%%%%%%%%%%%%%%%%%
%%%%%%%%%%%%%%%%%%%%%%%%%%%%%%%%%%%%%%%%%%%%%%%%%%%%%%%%%%%%%%%%%%%%%%
%%%%%%%%%%%%%%%%%%%%%%%%%%%%%%%%%%%%%%%%%%%%%%%%%%%%%%%%%%%%%%%%%%%%%%
%%%%%%%%%%%%%%%%%%%%%%%%%%%%%%%%%%%%%%%%%%%%%%%%%%%%%%%%%%%%%%%%%%%%%%

The first-order correction to the entropy we found in Eq.~(\ref{eq:entropy1}),
suggests that its value at all orders in $\alpha'$ is\footnote{It might seem
  that guessing the exact expression for the entropy just from the first order
  correction is too adventurous. However, to take this leap we profit from the
  results of the entropy function formalism at first and all orders, see
  Refs.~\cite{Sahoo:2006pm} and \cite{Sen:2007qy} respectively.}

 \begin{equation}
\label{eq:entropyexactfinal}
 \mathbb{S}_{(4)}
=
2\pi \sqrt{ N_{F1}N_{W} \left( n N_{S5}+2 \right)} \, .
\end{equation}

\noindent
Since the near-horizon solution preserves the symmetries of $\mathrm{AdS}_{3}
\times \mathrm{S}^{3}$, at the horizon $R_{(-)}\,^{a}\,_b=0$. Therefore all
corrections to the area law in the Wald entropy arise from the variation of
the Kalb-Ramond field strength with respect to the Riemann tensor.

Comparing our results with those on the literature, we observe an apparent
mismatch; previous studies based on near-horizon solutions obtain a correction
factor of value $+4$, instead of $+2$, inside the square root. The obvious
question is: why?

Before answering that question, let us notice that we can rewrite the
expression Eq.~(\ref{eq:entropyexactfinal}) substituting the number of
solitonic 5-branes $N_{S5}$ by the total, asymptotic solitonic 5-brane charge,
which we may call $\mathcal{Q}_{S5}$. In doing so, we get

\begin{equation}
 \mathbb{S}=2\pi \sqrt{ N_{F1}N_{W} \left( n \mathcal{Q}_{S5}+4-N_{\lambda} \right)} \, ,
\end{equation}

\noindent
where, we recall, $N_{\lambda}$ is the number of instantons in the
solution. Since previous studies do not include non-trivial gauge fields, we
see that setting $N_{\lambda}=0$ we reproduce the aforementioned correction
factor when the entropy is expressed in terms of total charges, instead of
fundamental constituents. As we are going to argue, this is no coincidence.

In the entropy function formalism, the S5 charge is magnetically carried by an
auxiliary Abelian vector whose Bianchi identity is \emph{uncorrected},
i.e. its field strength is a closed 2-form. This means that the S5 charge
carried by this auxiliary field is the same everywhere, asymptotically and at
the horizon. In most of the preceding literature this charge was identified
with $N_{S5}$, the number of S5-branes. On the other hand, it was observed in
Ref.~\cite{Prester:2010cw} that this charge is actually the asymptotic S5
charge magnetically carried by the Kalb-Ramond 2-form.

This identification is in direct contradiction with the identification of
total charges and sources which we have explained in full detail in the
preceding sections. As we have argued repeatedly, the $N_{S5}$ S5-branes are 
responsible of the local S5 charge measured at the horizon, while the
asymptotic charge results from adding to these the delocalized sources
introduced by the higher curvature corrections. Observe that, if the
asymptotic charge were $N_{S5}$, then the charge at the horizon would be given
by $(N_{S5}+2/n)$, with the shift caused by the higher curvature
corrections. It seems hard to justify how this would be possible, specially
taking into consideration that the curvature $R_{(-)}{}^{a}{}_{b}$ vanishes at
the horizon.

For all these reasons we claim that the correct expression for the entropy
expressed in terms of the number of fundamental constituents of the solution
is Eq.~(\ref{eq:entropyexactfinal}). This identification also matches the
microscopic result Eq.~(\ref{eq:entropyexact})
($k=n\mathcal{Q}_{S5}+2=nN_{S5}$).\footnote{Observe that the identification of
  $k$ in Ref.~\cite{Kutasov:1998zh} was made comparing with a solution of the
  zeroth-order in $\alpha'$ effective action.}

Notice that if we set $n=1$ the value of the entropy coincides with that of
the three-charge, five-dimensional black holes we described in
Ref.~\cite{Cano:2018qev}. From our perspective this seems completely natural,
since after setting $n=1$ the event horizon of the black hole we get is
identical to that of Ref.~\cite{Cano:2018qev}, so both Wald entropies should
match. Once again, this result would be in contradiction with the standard
expression found in the literature, where the correction factor is $+3$
instead of $+2$. Yet again, in this case, we find the origin of the
discrepancy might be on the identification of the charges. Notice that in the
5-dimensional black holes the hyperK\"ahler space is simply $\mathbb{R}^{4}$,
and the contribution to the asymptotic S5 charge coming from $\int
{R}_{(-)}{}^{{a}}{}_{{b}}\wedge{R}_{(-)}{}^{{b}}{}_{{a}}$ is just $-1$. Hence,
substituting $N_{S5}-1=\mathcal{Q}_{S5}$ in Eq.~(\ref{eq:entropyexactfinal}) we
reproduce the standard correction factor found in the literature.\footnote{At
  this stage, let us point out that the difference in the previously observed
  correction factors of $+3$ and $+4$ for these five- and four-dimensional
  black holes (sometimes referred as the \emph{$4D/5D$ connection}) had
  already been identified as caused by the negative unit of S5 charge carried
  by the Kaluza-Klein monopole of unit charge (for generic charge $n$, the
  associated S5 charge is $-1/n$), see \cite{Castro:2008ys,
    Castro:2008ne}. The relevance of the S5 charge carried by the KK monopole
  had already been remarked in Refs.~\cite{Sen:1997zb,Sen:1997js}.}

Therefore, in light of our results it seems that the appearance of a factor
different from $+2$ is caused by a misidentification of the number of
solitonic 5-branes of the heterotic superstring.

%%%%%%%%%%%%%%%%%%%%%%%%%%%%%%%%%%%%%%%%%%%%%%%%%%%%%%%%%%%%%%%%%%%%%%
%%%%%%%%%%%%%%%%%%%%%%%%%%%%%%%%%%%%%%%%%%%%%%%%%%%%%%%%%%%%%%%%%%%%%%
%%%%%%%%%%%%%%%%%%%%%%%%%%%%%%%%%%%%%%%%%%%%%%%%%%%%%%%%%%%%%%%%%%%%%%
%%%%%%%%%%%%%%%%%%%%%%%%%%%%%%%%%%%%%%%%%%%%%%%%%%%%%%%%%%%%%%%%%%%%%%
\section*{Acknowledgments}
%%%%%%%%%%%%%%%%%%%%%%%%%%%%%%%%%%%%%%%%%%%%%%%%%%%%%%%%%%%%%%%%%%%%%%
%%%%%%%%%%%%%%%%%%%%%%%%%%%%%%%%%%%%%%%%%%%%%%%%%%%%%%%%%%%%%%%%%%%%%%
%%%%%%%%%%%%%%%%%%%%%%%%%%%%%%%%%%%%%%%%%%%%%%%%%%%%%%%%%%%%%%%%%%%%%%
%%%%%%%%%%%%%%%%%%%%%%%%%%%%%%%%%%%%%%%%%%%%%%%%%%%%%%%%%%%%%%%%%%%%%%

We would like to thank A. Sen for useful correspondence. This work has been
supported in part by the MINECO/FEDER, UE grants FPA2015-66793-P and
FPA2015-63667-P, by the Italian INFN and by the Spanish Research Agency
(Agencia Estatal de Investigaci\'on) through the grant IFT Centro de
Excelencia Severo Ochoa SEV-2016-0597. AR is supported by ``Centro de
Excelencia Internacional UAM/CSIC'' and ``Residencia de Estudiantes''
fellowships. The work of PAC is funded by Fundaci\'on la Caixa through a ``la Caixa - Severo Ochoa''
International pre-doctoral grant. TO wishes to thank M.M.~Fern\'andez for her permanent support.

%%%%%%%%%%%%%%%%%%%%%%%%%%%%%%%%%%%%%%%%%%%%%%%%%%%%%%%%%%%%%%%%%%%%%%
%%%%%%%%%%%%%%%%%%%%%%%%%%%%%%%%%%%%%%%%%%%%%%%%%%%%%%%%%%%%%%%%%%%%%%
%%%%%%%%%%%%%%%%%%%%%%%%%%%%%%%%%%%%%%%%%%%%%%%%%%%%%%%%%%%%%%%%%%%%%%
%%%%%%%%%%%%%%%%%%%%%%%%%%%%%%%%%%%%%%%%%%%%%%%%%%%%%%%%%%%%%%%%%%%%%%
\appendix
%%%%%%%%%%%%%%%%%%%%%%%%%%%%%%%%%%%%%%%%%%%%%%%%%%%%%%%%%%%%%%%%%%%%%%
%%%%%%%%%%%%%%%%%%%%%%%%%%%%%%%%%%%%%%%%%%%%%%%%%%%%%%%%%%%%%%%%%%%%%%
%%%%%%%%%%%%%%%%%%%%%%%%%%%%%%%%%%%%%%%%%%%%%%%%%%%%%%%%%%%%%%%%%%%%%%
%%%%%%%%%%%%%%%%%%%%%%%%%%%%%%%%%%%%%%%%%%%%%%%%%%%%%%%%%%%%%%%%%%%%%%
\section{Instanton number density as a Laplacian}
\label{sec-instanton}
%%%%%%%%%%%%%%%%%%%%%%%%%%%%%%%%%%%%%%%%%%%%%%%%%%%%%%%%%%%%%%%%%%%%%%
%%%%%%%%%%%%%%%%%%%%%%%%%%%%%%%%%%%%%%%%%%%%%%%%%%%%%%%%%%%%%%%%%%%%%%
%%%%%%%%%%%%%%%%%%%%%%%%%%%%%%%%%%%%%%%%%%%%%%%%%%%%%%%%%%%%%%%%%%%%%%
%%%%%%%%%%%%%%%%%%%%%%%%%%%%%%%%%%%%%%%%%%%%%%%%%%%%%%%%%%%%%%%%%%%%%%

Consider an arbitrary Gibbons-Hawking space, with metric

\begin{equation}
 d\sigma^{2}=\mathcal{H}^{-1}(d\eta+\chi)^{2}+\mathcal{H} dx^{x} dx^{x}\,,
\qquad 
d\mathcal{H}= L\star_{(3)}d\chi\,,
\end{equation}

\noindent
and define on it an $\mathrm{SU}(2)$ field of the form

\begin{equation}
 A^{A}=- \mathcal{H}^{-1}\Phi^{A} (d\eta+\chi)+\breve{A}^{A}\,,
\end{equation}

\noindent
where $\Phi^{A}$ and $\breve{A}^{A}$ are, respectively, a function and a $1$-form
defined on $\mathbb{E}^{3}$.  Then the requirement of self-duality for the field
strength

\begin{equation}
 F^{A} = dA^{A}+\tfrac{1}{2} \epsilon^{ABC} A^{B}\wedge A^{C}
\end{equation}

\noindent
is equivalent to the Bogomol'nyi equation

\begin{equation}
F^{A}= +\star_{(4)} F^{A}\, ,
\qquad \Longleftrightarrow\qquad 
\mathcal{\breve D} \Phi^{A}=\star_{3} \breve F^{A}\, ,
\end{equation}

\noindent
and one has

\begin{align}
F^{A} 
& = 
-\mathcal{\breve D}\left( \mathcal{H}^{-1}\Phi^{A} \right)\wedge
(d\eta+\chi)-\mathcal{H}^{-1}\Phi^{A} d\chi + \breve F^{A}
\nonumber\\
\nonumber\\
& =
-\mathcal{\breve D}\left( \mathcal{H}^{-1}\Phi^{A} \right)\wedge
(d\eta+\chi)+\star_{3}\mathcal{\breve D}\left( \mathcal{H}^{-1}\Phi^{A}
\right)\, .
\end{align}

The instanton number density is, then.

\begin{align}
 F^{A}\wedge F^{A} 
& = -2\mathcal{\breve D}( \mathcal{H}^{-1}\Phi^{A}
)\wedge\star_{3}\mathcal{\breve D}( \mathcal{H}^{-1}\Phi^{A} ) \wedge
(d\eta+\chi)
\nonumber\\
\nonumber\\
& = 
-\mathcal{H}^{-1}\Big[ 2 \mathcal{\breve D}( \mathcal{H}^{-1}\Phi^{A} )
\wedge\star_{3}\mathcal{\breve D}\Phi^{A}
-2 \mathcal{H}^{-1}\Phi^{A}\mathcal{\breve D}( \mathcal{H}^{-1}\Phi^{A}
)\wedge\star_{3}\mathcal{\breve D}H
\nonumber\\
\nonumber\\
&
\phantom{=-\mathcal{H}^{-1}\Big[}+2 \mathcal{H}^{-1}\Phi^{A}\mathcal{\breve
  D}\star_{3}\mathcal{\breve D}\Phi^{A} -\mathcal{H}^{-2}\Phi^{A}\Phi^{A}
d\star_{3}d \mathcal{H} \Big]\wedge  (d\eta+\chi)
\nonumber\\
\nonumber\\
& =
-\mathcal{H}^{-1}\mathcal{\breve D}\star_{3}\mathcal{\breve D}\left(
  \mathcal{H}^{-1}\Phi^{A} \Phi^{A}\right)\wedge
(d\eta+\chi)=\mathcal{H}^{-1} \partial_{\underline{x}}\partial_{\underline{x}}\left(
  \mathcal{H}^{-1}\Phi^{A} \Phi^{A}\right)|v| d^{4}x
\nonumber \\
\nonumber \\
& = 
\nabla^{2}_{(4)}\left(\Phi^{A}\Phi^{A}/\mathcal{H}\right) |v| d^{4} x\, ,
\end{align}

\noindent
where we made use of the fact that $\mathcal{H}$ is harmonic, $d\star_{3}d \mathcal{H}=0$, and that $\mathcal{\breve D}\star_{3}\mathcal{\breve D}\Phi^{A} =\mathcal{\breve D}\breve F^{A}=0  $.

%If $F^{A_{i}}= +\star_{(4)} F^{A_{i}}$

%%%%%%%%%%%%%%%%%%%%%%%%%%%%%%%%%%%%%%%%%%%%%%%%%%%%%%%%%%%%%%%%%%%%%%
%%%%%%%%%%%%%%%%%%%%%%%%%%%%%%%%%%%%%%%%%%%%%%%%%%%%%%%%%%%%%%%%%%%%%%
%%%%%%%%%%%%%%%%%%%%%%%%%%%%%%%%%%%%%%%%%%%%%%%%%%%%%%%%%%%%%%%%%%%%%%
%%%%%%%%%%%%%%%%%%%%%%%%%%%%%%%%%%%%%%%%%%%%%%%%%%%%%%%%%%%%%%%%%%%%%%
\section{Regular instantons over Kaluza-Klein monopoles}
\label{sec:regular_inst}
%%%%%%%%%%%%%%%%%%%%%%%%%%%%%%%%%%%%%%%%%%%%%%%%%%%%%%%%%%%%%%%%%%%%%%
%%%%%%%%%%%%%%%%%%%%%%%%%%%%%%%%%%%%%%%%%%%%%%%%%%%%%%%%%%%%%%%%%%%%%%
%%%%%%%%%%%%%%%%%%%%%%%%%%%%%%%%%%%%%%%%%%%%%%%%%%%%%%%%%%%%%%%%%%%%%%
%%%%%%%%%%%%%%%%%%%%%%%%%%%%%%%%%%%%%%%%%%%%%%%%%%%%%%%%%%%%%%%%%%%%%%

Let us consider the monopoles introduced in Section~\ref{Instantons}.  We are
going to see that, over the KK monopole of unit charge $(n=1)$, all these
monopoles give rise to regular instantons, giving an explicit construction of
the instanton bundles. These instantons over a KK monopole with unit charge
were first described in Ref.~\cite{BoutalebJoutei:1979iz}. Moreover, for
higher KK monopole charge, $n>1$, the corresponding instantons are also
regular in the full spacetime metric provided the black hole horizon has
non-vanishing area.  From now on we consider the $n=1$ case in detail for the
sake of simplicity. The regularity condition for the $n>1$ case is treated in
Section~\ref{sec-mu}.

WE are going to use the following convenient way of writing the gauge fields
$A^{A}$ in Eq.~(\ref{vectorAA}) in the case $n=1$ 

\begin{eqnarray}
A^{A}
& = &
-h(r)r^{2} v^{A}_{R}
+\frac{x^{A}}{r}(d\Psi+d\phi \cos{\theta})\left(\frac{R}{2\mathcal{H}} r f(r)
-h(r)r^{2}\right)\, ,
\end{eqnarray}

\noindent
where $v^{A}_{R}$ are the right-invariant Maurer-Cartan forms

\begin{equation}
\begin{cases}
v^{1}_{R}=\sin\phi d\theta-\sin{\theta}\cos\Psi d\Psi\, ,\\
v^{2}_{R}=\cos\phi d\theta+\sin{\theta}\sin\phi d\Psi\, ,\\
v^{3}_{R}=d\phi+\cos{\theta} d\Psi\, ,\\
\end{cases}
\hspace{1cm}  
dv^{A}_{R}+ \tfrac{1}{2}\epsilon_{ABC}v^{B}_{R}\wedge v^{C}_{R}=0\, .
\end{equation}

%%%%%%%%%%%%%%%%%%%%%%%%%%%%%%%%%%%%%%%%%%%%%%%%%%%%%%%%%%%%%%%%%%%%%%
%%%%%%%%%%%%%%%%%%%%%%%%%%%%%%%%%%%%%%%%%%%%%%%%%%%%%%%%%%%%%%%%%%%%%%
%%%%%%%%%%%%%%%%%%%%%%%%%%%%%%%%%%%%%%%%%%%%%%%%%%%%%%%%%%%%%%%%%%%%%%
%%%%%%%%%%%%%%%%%%%%%%%%%%%%%%%%%%%%%%%%%%%%%%%%%%%%%%%%%%%%%%%%%%%%%%
\subsection{Near-origin limit}
%%%%%%%%%%%%%%%%%%%%%%%%%%%%%%%%%%%%%%%%%%%%%%%%%%%%%%%%%%%%%%%%%%%%%%
%%%%%%%%%%%%%%%%%%%%%%%%%%%%%%%%%%%%%%%%%%%%%%%%%%%%%%%%%%%%%%%%%%%%%%
%%%%%%%%%%%%%%%%%%%%%%%%%%%%%%%%%%%%%%%%%%%%%%%%%%%%%%%%%%%%%%%%%%%%%%
%%%%%%%%%%%%%%%%%%%%%%%%%%%%%%%%%%%%%%%%%%%%%%%%%%%%%%%%%%%%%%%%%%%%%%

When $r\to 0$, $\mathcal{H}\sim (R/2)/r$, and after the change of variables
$r=\frac{\rho^{2}}{2 R}$, we can rewrite the metric as

\begin{equation}
d\sigma^{2}
=
d\rho^{2}+\rho^{2}d\Omega^{2}_{(3)}\, ,
\end{equation}

\noindent
where

\begin{equation}
d\Omega^{2}_{(3)}
=
\tfrac{1}{4}
\left[(d\Psi+d\phi\cos{\theta})^{2}+d\Omega^{2}_{(2)}\right]\, ,  
\end{equation}

\noindent
is the metric of the round $\mathrm{S}^{3}$ of unit radius, so the space is locally
$\mathbb{E}^{4}$. 

On the other hand, at the origin $r=\rho=0$, the gauge fields of the different
instantons take the values

\begin{equation}
\lim_{r\rightarrow 0}A_{\mu,s}
=
\left\{
  \begin{array}{l}
{\displaystyle\frac{1}{g}}v_{R}\, ,\,\,\,\, s\neq 0\, , \\
\\
0\, ,\,\,\,\,s=0\, ,\\
\end{array}
\right. \quad
\lim_{r\rightarrow 0}A_{\lambda}=\displaystyle\frac{1}{g}v_{R}
\end{equation}

\noindent
where $A=A^{A}T_{A}$ and $v_{R}=v^{A}_{R}T_{A}$ and $\{T_{A}\}$ is the basis
of the $\mathfrak{su}(2)$ algebra

\begin{equation}
T_{A}
=
-\tfrac{i}{2}\sigma^{A}\, ,
\quad 
\left[T_{A},T_{B}\right]=+\epsilon_{ABC}T_{C}\, .
\end{equation}

Hence, except in the case $(\mu,s=0)$, these vectors are singular at the
origin, something which can be made explicit if we write $A(r=0)$ in terms of
the Vielbein basis. However, $A(r=0)$ is pure gauge and the field strength $F$
is finite at the origin for $s\neq 0$. Furthermore, Kronheimer's conditions
are met there. This indicates the existence, for $s\neq 0$, of another gauge in
which the gauge field is regular at the origin.

An arbitrary gauge transformation of the vector reads

\begin{equation}
\label{eq:gengauge}
\hat{A}=V A V^{-1}-\frac{1}{g}d V V^{-1\, }\, ,
\end{equation}

\noindent
where $V\in \mathrm{SU}(2)$. If we choose $V=U$ where

\begin{equation}
\label{Utransf}
U
\equiv
e^{-T^{3}\Psi}e^{-T^{2}\theta}e^{-T^{1}\phi}\,,
\end{equation}

\noindent
is the generic $\mathrm{SU}(2)$ group element parametrized in terms of the
Euler angles $\Psi,\theta,\phi$, and we take into account the properties of the
Maurer-Cartan forms\footnote{These properties follow from their definition in
  terms of $U$:
\begin{equation}
v_{L}\equiv -U^{-1}dU\, ,
\quad 
v_{R}\equiv -dU U^{-1}\, .
\end{equation}
}

\begin{equation}
v_{R} = Uv_{L}U^{-1}\, ,
\end{equation}

\noindent
we get the following transformed gauge field:\footnote{
The left-invariant MC forms read
\begin{equation}
\begin{cases}
v^{1}_{L}=-\sin\Psi d\theta+\sin{\theta}\cos\Psi d\phi\, \\
v^{2}_{L}=\cos\Psi d\theta+\sin{\theta}\sin\Psi d\phi\, \\
v^{3}_{L}=d\Psi+\cos{\theta} d\phi\, \\
\end{cases}
\hspace{1cm}  
dv^{A}_{L}- \tfrac{1}{2}\epsilon_{ABC}v^{B}_{L}\wedge v^{C}_{L}=0\, .
\end{equation}}

\begin{eqnarray}
g\hat{A}
& = &
-(1+gh(r)r^{2} )v_{L}
-g\left(\frac{R}{2\mathcal{H}} r f(r)-h(r)r^{2}\right)
(d\Psi+\cos{\theta} d\phi)T_{3}\, .
\end{eqnarray}

\noindent
If we apply this transformation to the $\lambda$ and $(\mu,s>0)$-cases, we get,
respectively:

\begin{eqnarray}\label{Alambda}
g\hat{A}_{\lambda}
& = &
-\frac{\lambda^{2} r}{1+\lambda^{2} r}\left(v^{1}_{L}T_{1}+v^{2}_{L}T_{2}\right)
-\frac{r\left(1+\lambda^{2}(r+R/2)\right)}{(R/2+r)(1+\lambda^{2} r)}v^{3}_{L}T_{3}\, ,  
\\
& & \nonumber \\
\label{Amus}
g\hat{A}_{\mu,s>0}
& = & 
-\frac{\mu r}{\sinh(\mu r+s)}\left(v^{1}_{L}T_{1}+v^{2}_{L}T_{2}\right)
-\frac{r\left[1+\mu R/2 \coth(\mu r+s)\right]}{R/2+r}v^{3}_{L}T_{3}\, .
\end{eqnarray}

Now, near the origin the vectors behave as $\hat{A}\sim\rho^{2} v_{L}$ and we conclude 
that the gauge fields are regular at $r=0$ for all values of $\lambda$ and $s>0$. 
In the case $s=0$ the vector is already regular and we do not need to perform any
gauge transformation. In that case, it reads 

\begin{multline}
\label{Amu0}
g\hat{A}_{\mu,s=0}
 = 
\left(1-\frac{\mu r}{\sinh{\mu r}}\right)v_{R}  
\\
\\
+\frac{x^{A}}{r}T_{A}(d\Psi+d\phi\cos{\theta})\frac{r}{r+R/2}
\left\{
1-\frac{\mu\left[r+R/2(1-\cosh{\mu r})\right]}{\sinh{\mu r}}
\right\}\, .
\end{multline}

%%%%%%%%%%%%%%%%%%%%%%%%%%%%%%%%%%%%%%%%%%%%%%%%%%%%%%%%%%%%%%%%%%%%%%
%%%%%%%%%%%%%%%%%%%%%%%%%%%%%%%%%%%%%%%%%%%%%%%%%%%%%%%%%%%%%%%%%%%%%%
%%%%%%%%%%%%%%%%%%%%%%%%%%%%%%%%%%%%%%%%%%%%%%%%%%%%%%%%%%%%%%%%%%%%%%
%%%%%%%%%%%%%%%%%%%%%%%%%%%%%%%%%%%%%%%%%%%%%%%%%%%%%%%%%%%%%%%%%%%%%%
\subsection{Asymptotic limit}
%%%%%%%%%%%%%%%%%%%%%%%%%%%%%%%%%%%%%%%%%%%%%%%%%%%%%%%%%%%%%%%%%%%%%%
%%%%%%%%%%%%%%%%%%%%%%%%%%%%%%%%%%%%%%%%%%%%%%%%%%%%%%%%%%%%%%%%%%%%%%
%%%%%%%%%%%%%%%%%%%%%%%%%%%%%%%%%%%%%%%%%%%%%%%%%%%%%%%%%%%%%%%%%%%%%%
%%%%%%%%%%%%%%%%%%%%%%%%%%%%%%%%%%%%%%%%%%%%%%%%%%%%%%%%%%%%%%%%%%%%%%

In order to study the asymptotic limit, it is convenient to recall the global
structure of the solution and to use cartesian coordinates $x^{1}, x^{2},
x^{3}$.  In these coordinates, the two 1-forms $\chi^{(\pm)}$
read\footnote{For convenience, in this section $\chi$ is defined locally as
  $\chi=d\phi \cos{\theta}$, without the factor of $q$ that appears in
  (\ref{chi-1}).}

\begin{equation}
\chi^{(\pm)}
=
\frac{x^{1}dx^{2}-x^{2} dx^{1}}{r(x^{3}\pm r)}\, ,
\quad \text{where}\,\,\, 
r\equiv \sqrt{(x^{1})^{2}+(x^{2})^{2}+(x^{3})^{2}}\, .
\end{equation}

\noindent
We use $\chi^{(+)}$ in the upper space $x^{3}\ge 0$ and $\chi^{(-)}$ in the
lower one $x^{3}\le 0$ so that $\chi^{(\pm)}$ are regular in their respective
regions. Moreover, we observe that

\begin{equation}\label{chiinf}
\lim_{r\rightarrow\infty}\chi^{(\pm)}=0\, ,
\end{equation}

\noindent
where the limit is again taken in the respective region.  In this limit we
also have $\mathcal{H}\sim 1$, and hence the space becomes the direct product
$\mathbb{E}^{3}\times \mathrm{S}^{1}$. The metric takes the form

\begin{equation}
d\sigma^{2}
=
\frac{R^{2}}{4}d\Psi^{2}+dx^{i}dx^{i}\, .
\end{equation}

Let us now explore the asymptotic behavior of the gauge fields. 

In the case of $\hat{A}_{\lambda}$ we obtain

\begin{equation}
\lim_{r\rightarrow\infty} \hat{A}_{\lambda}
=
\lim_{r\rightarrow\infty}\frac{-1}{g}v_{L}
=-\frac{1}{g}T_{3}d\Psi\, .
\end{equation}

\noindent
The first equality is obtained by using the explicit dependence in $r$ in
Eq.~(\ref{Alambda}), while in the second equality we use the implicit dependence
contained in the angular 1-forms inside the MC forms.\footnote{For example,
  $\lim_{r\rightarrow\infty} d\theta=0$, which is clear if we take into
  account that the norm of $d\theta$ vanishes: $|d\theta|^{2}= r^{-2}$ when
  $r\rightarrow\infty$.  On the other hand, the norm of $d\Psi$ remains finite
  because it is a compact dimension: $|d\Psi|^{2}\rightarrow 4/R^{2}$.}  The
vector at infinity is pure gauge, so that there is an element
$V\in\mathrm{SU}(2)$ such that

\begin{equation}
\hat{A}_{\lambda\, \infty}=-\frac{1}{g}dVV^{-1}\, .
\end{equation}

\noindent
This element is

\begin{equation}
V
=
e^{\Psi T_{3}}
=
\begin{pmatrix}
e^{-i\Psi/2} & 0                \\
0                 & e^{i\Psi/2} \\
\end{pmatrix},
\end{equation}

\noindent
which defines a map from the equator of the sphere to the subgroup
$\mathrm{U}(1)\subset \mathrm{SU}(2)$. Hence, $V$ is an homomorphism from the
circle in itself which belongs to the first homotopy class.

This case is very simple because the asymptotic value of the vector does not
depend on free parameters. The family $(\mu,s)$ is more interesting in this
sense.  In the case $s>0$, we obtain from Eq.~(\ref{Amus})

\begin{equation}
\lim_{r\rightarrow\infty} \hat{A}_{\mu,s}
=
\lim_{r\rightarrow\infty}\frac{-(\mu R/2+1)}{g}(d\Psi+\chi)T_{3}
=
-\frac{(\mu R/2+1)}{g}T_{3}d\Psi\, ,
\end{equation}

\noindent
where we used that $v^{3}_{L}=(d\Psi+\chi)$ and Eq.~(\ref{chiinf}). Therefore,
the vector at infinity is pure gauge, and the corresponding gauge
transformation reads
\begin{equation}
V
=
e^{(\mu R/2+1)\Psi T_{3}}
=
\begin{pmatrix}
e^{-i(\mu R/2+1)\Psi/2} & 0                \\
0                 & e^{i(\mu R/2+1)\Psi/2} \\
\end{pmatrix}\, .
\end{equation}

\noindent
However, we must demand that $V$ is a single-valued map
$V:\mathrm{S}^{1}\subset\mathrm{S}^{3}\rightarrow \mathrm{U}(1)\subset
\mathrm{SU}(2)$.  Since $\Psi$ has period $4\pi$ we see that the following
quantization condition must hold so that $V(\Psi+4\pi)=V(\Psi)$:

\begin{equation}
\frac{\mu R}{2}+1\in \mathbb{Z}\, .
\end{equation}

The case $s=0$ can be reduced to the previous one after we apply the gauge
transformation $U^{-1}$ in (\ref{Utransf}) to the asymptotic value of
Eq.~(\ref{Amu0}). Since we demand that $\mu\ge0$, we can write

\begin{equation}
\label{mmu}
\mu=\frac{2m}{R}, \quad m=0,1,2,\ldots
\end{equation}

%%%%%%%%%%%%%%%%%%%%%%%%%%%%%%%%%%%%%%%%%%%%%%%%%%%%%%%%%%%%%%%%%%%%%%
%%%%%%%%%%%%%%%%%%%%%%%%%%%%%%%%%%%%%%%%%%%%%%%%%%%%%%%%%%%%%%%%%%%%%%
%%%%%%%%%%%%%%%%%%%%%%%%%%%%%%%%%%%%%%%%%%%%%%%%%%%%%%%%%%%%%%%%%%%%%%
%%%%%%%%%%%%%%%%%%%%%%%%%%%%%%%%%%%%%%%%%%%%%%%%%%%%%%%%%%%%%%%%%%%%%%
\subsection{Higher KK monopole charge with a horizon}
\label{sec-mu}
%%%%%%%%%%%%%%%%%%%%%%%%%%%%%%%%%%%%%%%%%%%%%%%%%%%%%%%%%%%%%%%%%%%%%%
%%%%%%%%%%%%%%%%%%%%%%%%%%%%%%%%%%%%%%%%%%%%%%%%%%%%%%%%%%%%%%%%%%%%%%
%%%%%%%%%%%%%%%%%%%%%%%%%%%%%%%%%%%%%%%%%%%%%%%%%%%%%%%%%%%%%%%%%%%%%%
%%%%%%%%%%%%%%%%%%%%%%%%%%%%%%%%%%%%%%%%%%%%%%%%%%%%%%%%%%%%%%%%%%%%%%

When the charge of the KK monopole is bigger than one, the 4-dimensional
manifold it describes contains an conical singularity at the origin due to a
deficit in the angle covered by the coordinate $\Psi$. This property also
affects the instanton field; see, \textit{e.g.}, the appearance of $n$ in
Eq.~(\ref{vectorAA}). Because of this factor, we have not been able to mimic
the steps of the previous sections and find a gauge transformation rendering
the fields zero at the origin, which seem to remain singular at this location
of this GH space.

That the instantons are singular can be seen from the fact that the
angular coordinates are ill-defined at $r=0$, while in this basis some of the
angular components of corresponding vectors 1-form are non-vanishing. However
this problem is immediately solved in the black hole solutions that we
consider if the event horizon has non-vanishing area. There, the angular
coordinates are perfectly valid at $r=0$ and the gauge fields are already
regular in the original gauge (at this location the coordinate $r$ is not well
defined, but the corresponding component of the 1-forms of all the instantons
considered is zero).

Having said that, it is easy to check that the instantons of the
$\lambda$-family are globally regular, since they vanish asymptotically. On
the other hand, for the $\mu,s$-family we obtain the following asymptotic
behavior

\begin{equation}
\lim_{r\rightarrow\infty} g A_{\mu,s}
=
\frac{\mu R}{2} (d \Psi+ n \cos \theta d\phi) \frac{x^{A}}{r}T_{A} \, ,
\end{equation}

\noindent
with $\chi=q \cos{\theta} d\phi$. Written in this manner, we see that the gauge field contains a string singularity at $\theta=0,\pi$ that extends to infinity, while remaining regular elsewhere. The local expression around these locations reduces to

\begin{equation}
\lim_{\substack{r\rightarrow\infty\\ \theta\rightarrow 0,\pi}} g A_{\mu,s}=\mp \frac{\mu R}{2} (d \Psi \pm n d\phi) T_{3} \, ,
\end{equation}

\noindent
for the upper and lower plane respectively.

We now perform a local gauge transformation \eqref{eq:gengauge} in a small
open set around these two points with

\begin{equation}
\label{eq:gtmu}
V^{(\pm)}=e^{\mp \mu n R\phi T_{3}/2}= 
\begin{pmatrix}
e^{\pm i  \mu n R\phi/4} & 0                \\
0                 & e^{\mp i { \mu n R\phi}/{4}} \\
\end{pmatrix} \, ,
\end{equation}

\noindent
and get

\begin{equation}
\lim_{r\rightarrow\infty} g \hat{A}_{\mu,s} 
= 
\mp \left(\frac{\mu R}{2} d \Psi\right) T_{3} \, ,
\end{equation}

\noindent 
which have no string singularity whatsoever. As shown in
Ref.~\cite{BoutalebJoutei:1979iy}, the local gauge transformations
Eq.~(\ref{eq:gtmu}) can be extended through the sphere (except the poles) by
replacing $T_{3} \rightarrow x^{A} T_{A} /r$. Then, in the intersections
$\hat{A}^{(+)}$ and $\hat{A}^{(-)}$ are related by the gauge transformation

\begin{equation}
V^{(\pm)}=e^{\mp \mu n R \phi  (x^{A}/r) T_{A}} \, ,
\end{equation}

\noindent
which is single valued in the coordinate $\phi$ if and only if $\mu n R /2$ is
an integer number. Therefore, we get a generalization of the quantization
condition Eq.~(\ref{mmu}) for the $n>1$ case,

\begin{equation}
\label{mmun}
\mu=\frac{2m}{R n}\, , 
\quad 
m=0,1,2,\ldots
\end{equation}

%%%%%%%%%%%%%%%%%%%%%%%%%%%%%%%%%%%%%%%%%%%%%%%%%%%%%%%%%%%%%%%%%%%%%%
%%%%%%%%%%%%%%%%%%%%%%%%%%%%%%%%%%%%%%%%%%%%%%%%%%%%%%%%%%%%%%%%%%%%%%
%%%%%%%%%%%%%%%%%%%%%%%%%%%%%%%%%%%%%%%%%%%%%%%%%%%%%%%%%%%%%%%%%%%%%%
%%%%%%%%%%%%%%%%%%%%%%%%%%%%%%%%%%%%%%%%%%%%%%%%%%%%%%%%%%%%%%%%%%%%%%
\subsection{Contribution to the solution}
%%%%%%%%%%%%%%%%%%%%%%%%%%%%%%%%%%%%%%%%%%%%%%%%%%%%%%%%%%%%%%%%%%%%%%
%%%%%%%%%%%%%%%%%%%%%%%%%%%%%%%%%%%%%%%%%%%%%%%%%%%%%%%%%%%%%%%%%%%%%%
%%%%%%%%%%%%%%%%%%%%%%%%%%%%%%%%%%%%%%%%%%%%%%%%%%%%%%%%%%%%%%%%%%%%%%
%%%%%%%%%%%%%%%%%%%%%%%%%%%%%%%%%%%%%%%%%%%%%%%%%%%%%%%%%%%%%%%%%%%%%%

These instantons contribute to the solution through the function $\mathcal{Z}_{0}$. 
Using the general result Eq.~(\ref{eq:Z0}), their contributions read

\begin{eqnarray}
\label{eq:instZ0a}
\Delta \mathcal{Z}_{0}\Big|_{\lambda}
& = & 
-\frac{2\alpha'}{r(r+q)(\lambda^{2} r+1)^{2}}\, ,
\\
& & \nonumber \\
\label{eq:instZ0b}
\Delta \mathcal{Z}_{0}\Big|_{\mu,s>0}
& = & 
-2\alpha' \frac{\left[1-\mu r\coth{\left(\mu r+ s\right)}\right]^{2}}{r(r+q)}\,
,
\\
& & \nonumber \\
\label{eq:instZ0c}
\Delta \mathcal{Z}_{0}\Big|_{\mu,s=0}
& = & 
-2\alpha' \frac{\left[1-\mu r\coth{\mu r}\right]^{2}}{r(r+q)} \, ,
\end{eqnarray}

\noindent
where $q=nR/2$ and we set the gauge coupling constant in the Heterotic
Superstring to $g=1$.

Now the pole and asymptotic constant can be removed by adding an appropriate
harmonic function $a+c/r$.  After this transformation, the contributions read
 
\begin{eqnarray}
\Delta \mathcal{Z}_{0}\Big|_{\lambda}
& = & 
2\alpha'\frac{(r+q)\lambda^{2}(\lambda^{2} r+2)+1}{q(r+q)(\lambda^{2}
  r+1)^{2}}\, ,
\\
& & \nonumber \\
\Delta \mathcal{Z}_{0}\Big|_{\mu,s>0}
& = &
2\alpha' \left\{\mu^{2}+\frac{q^{-1}}{r}
-\frac{\left[1-\mu r\coth{(\mu r+ s)}\right]^{2}}{r(r+q)}\right\} \, ,
\\
& & \nonumber \\
\Delta \mathcal{Z}_{0}\Big|_{\mu,s=0}
& = & 
2\alpha' \left\{\mu^{2}
-\frac{\left[1-\mu r\coth{\mu r}\right]^{2}}{r(r+q)}\right\} \, .
\end{eqnarray}

%%%%%%%%%%%%%%%%%%%%%%%%%%%%%%%%%%%%%%%%%%%%%%%%%%%%%%%%%%%%%%%%%%%%%%
%%%%%%%%%%%%%%%%%%%%%%%%%%%%%%%%%%%%%%%%%%%%%%%%%%%%%%%%%%%%%%%%%%%%%%
%%%%%%%%%%%%%%%%%%%%%%%%%%%%%%%%%%%%%%%%%%%%%%%%%%%%%%%%%%%%%%%%%%%%%%
%%%%%%%%%%%%%%%%%%%%%%%%%%%%%%%%%%%%%%%%%%%%%%%%%%%%%%%%%%%%%%%%%%%%%%
%%%%%%%%%%%%%%%%%%%%%%%%%%%%%%%%%%%%%%%%%%%%%%%%%%%%%%%%%%%%%%%%%%%%%%
%%%%%%%%%%%%%%%%%%%%%%%%%%%%%%%%%%%%%%%%%%%%%%%%%%%%%%%%%%%%%%%%%%%%%%

\end{document}